\documentclass[12pt,preprint]{aastex}
\usepackage{amssymb, amsmath, apjfonts, emulateapj5, natbib, color}
\input psfig.tex

\def\aa{{A\&A}}
\def\aas{{ A\&AS}}
\def\aj{{AJ}}

\def\amin{$^\prime$}
\def\annrev{{ARA\&A}}
\def\apj{{ApJ}}
\def\apjs{{ApJS}}
\def\asec{$^{\prime\prime}$}

\def\deg{$^{\circ}$}

\def\e#1{$\times$10$^{#1}$}

\def\kms{km s$^{-1}$}

\def\lax{{$\mathrel{\hbox{\rlap{\hbox{\lower4pt\hbox{$\sim$}}}\hbox{$<$}}}$}}
\def\gax{{$\mathrel{\hbox{\rlap{\hbox{\lower4pt\hbox{$\sim$}}}\hbox{$>$}}}$}}
\def\simlt{\lower.5ex\hbox{$\; \buildrel < \over \sim \;$}}
\def\simgt{\lower.5ex\hbox{$\; \buildrel > \over \sim \;$}}

\def\micron{{$\mu$m}}
\def\mnras{{MNRAS}}

\def\pasp{{PASP}}

\def\percm2{cm$^{-2}$}

\def\sb{mag~arcsec$^{-2}$}

\def\hi{\ion{H}{1}}

\slugcomment{To appear in {\it The Astrophysical Journal Supplement}.}
\shortauthors{HO ET AL.}

\begin{document}

\title{THE CARNEGIE-IRVINE GALAXY SURVEY. I. OVERVIEW AND ATLAS OF OPTICAL IMAGES}

\author{Luis C. Ho\altaffilmark{1}, Zhao-Yu Li\altaffilmark{1,2}, Aaron J. 
Barth\altaffilmark{3}, Marc S. Seigar\altaffilmark{4,5}, and Chien Y. 
Peng\altaffilmark{1,6}}

\altaffiltext{1}{The Observatories of the Carnegie Institution for Science, 
813 Santa Barbara Street, Pasadena, CA 91101, USA}

\altaffiltext{2}{Department of Astronomy, School of Physics, Peking 
University, Beijing 100871, China}

\altaffiltext{3}{Department of Physics and Astronomy, 4129 Frederick Reines 
Hall, University of California, Irvine, CA 92697-4575, USA}

\altaffiltext{4}{Department of Physics \& Astronomy, University of Arkansas 
at Little Rock, 2801 S. University Avenue, Little Rock, AR 72204, USA}

\altaffiltext{5}{Arkansas Center for Space and Planetary Sciences, 202 Old Museum Building, University of Arkansas, Fayetteville, AR 72701, USA}

\altaffiltext{6}{NRC Herzberg Institute of Astrophysics, 5071 West Saanich 
Road, Victoria, BC V9E 2E7, Canada}

\begin{abstract}
The  Carnegie-Irvine Galaxy Survey (CGS) is a long-term program to investigate 
the photometric and spectroscopic properties of a statistically complete 
sample of 605 bright ($B_T < 12.9$ mag), southern ($\delta < 0$\deg) galaxies 
using the facilities at Las Campanas Observatory.  This paper, the first in a 
series, outlines the scientific motivation of CGS, defines the sample, and 
describes the technical aspects of the optical broadband (\emph{BVRI}) imaging 
component of the survey, including details of the observing program, data 
reduction procedures, and calibration strategy.  The overall quality of the 
images is quite high, in terms of resolution (median seeing $\sim 
1$\asec), field of view (8\farcm9$\times$8\farcm9), and depth (median limiting 
surface brightness $\sim 27.5$, 26.9, 26.4, and 25.3 mag~arcsec$^{-2}$ in the 
$B$, $V$, $R$, and $I$ bands, respectively). We prepare a digital image atlas
showing several different renditions of the data, including three-color 
composites, star-cleaned images, stacked images to enhance faint features, 
structure maps to highlight small-scale features, and color index maps 
suitable for studying the spatial variation of stellar content and dust.  In 
anticipation of upcoming science analyses, we tabulate an extensive set of 
global properties for the galaxy sample.  These include optical isophotal and 
photometric parameters derived from CGS itself, as 
well as published information on multiwavelength (ultraviolet, $U$-band, 
near-infrared, far-infrared) photometry, internal kinematics (central stellar 
velocity dispersions, disk rotational velocities), environment (distance to 
nearest neighbor, tidal parameter, group or cluster membership), and \hi\ 
content.  The digital images and science-level data products will be made 
publicly accessible to the community.
\end{abstract}


\section{Motivation}

The structural components of a galaxy bear witness to the major episodes that
have shaped them during its life cycle, and, as such, provide crucial
fossil records of the physical processes operating in galaxy formation and 
evolution.  Morphological clues have long guided our intuition about galaxy 
formation \citep{Gott77, Wyse97}.  The two most conspicuous luminous 
components modulated along the Hubble sequence---the bulge and the disk---have
been the main focal points of our modern concepts of how galaxies were 
assembled.  The roughly spheroidal shape of elliptical galaxies and the bulges 
of spiral galaxies, along with the recognition of their generally evolved 
stellar population, signify rapid, dissipationless collapse at an early 
epoch.  The $r^{1/4}$ profile \citep{deVaucouleurs48} of ellipticals and 
classical bulges is often interpreted as a signature of viole0nt relaxation 
\citep{vanAlbada82} resulting from rapid assembly through major mergers.   By 
contrast, the flattened configuration of an exponential disk \citep{Freeman70}, 
along with their younger, more mixed stellar populations, suggests that more 
gradual, dissipative processes have been operating and are still ongoing today.

With the advent of modern, large-format detectors and the accompanying
improvement in linearity, dynamic range, and image resolution, our view of 
galaxy morphology has grown steadily more elaborate, to the point that, in many 
instances, the classical picture of a bulge plus disk no longer suffices to 
describe the complex details seen in state-of-the-art galaxy images.  While an 
$r^{1/4}$ law still provides a good first-order approximation to the overall 
light distribution of many elliptical galaxies, at least in images taken with 
typical ground-based resolutions\footnote{In detail, the global profiles of 
ellipticals span a wider range of shapes (see \citealt{Kormendy09} for a 
recent, comprehensive review).  At sub-arcsecond resolution, for instance, as 
afforded by the {\it Hubble Space Telescope}, the central light distributions 
show significant additional deviations from the global, outer profiles 
\citep[e.g.,][]{Lauer95, Ravindranath01}.}, the situation is considerably 
more complicated for the bulges of disk galaxies.  The central light 
distribution of not only spirals, but also S0s, shows a variety of shapes 
\citep[e.g.,][]{Andredakis94, Courteau96b, deJong96, MacArthur03, Laurikainen05, 
Graham08, Gadotti09}, which are often well represented by a \citet{Sersic68} 
$r^{1/n}$ 
profile, with $n$ ranging from 1 (pure exponential) to 4 (standard 
de~Vaucouleurs value).  Imaging with the {\it Hubble Space Telescope}\ reveals 
an even greater degree of structural heterogeneity, including nuclear disks, 
nuclear spirals and rings, central nuclei and star clusters, and intricate 
dust lanes \citep[e.g.,][]{Carollo97, Boker02, Seigar02}.

This rich variety of kinematically cold structures in the central regions of
galaxies compels us to reevaluate the very definition of a ``bulge.''  
Evidently many bulges are not the old, dead, fully established systems we once 
thought. Instead, they appear to have experienced a much more gradual, 
protracted history of formation in which secular evolutionary processes have 
been---and still are---at work.  While most of the recent attention has focused
on the phenomenon of ``pseudobulges'' (\citealt{Kormendy04}, and 
references therein), there has also been a growing appreciation that even the
classical bulges may have experienced a rather dynamic evolutionary history.
Apart from the prevalence of features such as kinematically distinct or
counterrotating cores, which are indicative of discrete accretion events
\citep[e.g.,][]{Forbes95}, many early-type galaxies contain nuclear gaseous
disks and dust lanes \citep[e.g.,][]{vanDokkum95, Tran01} and 
central stellar disks \citep[e.g.,][]{Rix92, Ledo10}, 
concrete reminders that these are continually evolving systems.  In addition, 
a large fraction of disk galaxies \citep[$\sim 30$\%;][]{Lutticke00}, 
including S0s \citep{Aguerri05}, possess boxy or peanut-shaped bulges,
which are believed to be not actual bulges at all but edge-on bars 
\citep[e.g.,][]{Bureau99, Athanassoula05, Kormendy10}.  This 
confounds our usual notion of what a bulge is, even in galaxies that are 
traditionally viewed as bulge-dominated.

The bulge does not live in isolation but is intimately linked with an extended
disk, whose prominence relative to the bulge varies along the Hubble sequence.
While to first order the disk can be approximated by a single exponential 
profile, which arises as a natural consequence of viscous dissipation (e.g., 
\citealt{Lin87}), in detail it, too, exhibits a plethora of morphologically
distinct features that may provide insights into more global formation 
processes.  Pronounced departures occur at both small and large radii.  At small
radii, typically over scales where the bulge begins to dominate, some 
disks exhibit a downturn compared to the inner extrapolation of the outer 
exponential profile; \citet{Freeman70} called these ``Type II'' profiles.  The 
central regions of late-type spirals, on the other hand, often show excesses 
above the outer extrapolated disk (\citealt{Boker03}).  On large scales, 
\citet{vanderKruit79} first noticed that the disks of some spiral galaxies have 
a sharply truncated outer edge, as opposed to those that maintain a single 
exponential profile that merges smoothly with the sky background.  By contrast,
there is a growing population of disks that shows the opposite effect---an 
upturn rather than a downturn---at large radii (\citealt{Erwin05}).  Still 
others possess what appears to be a distinct, highly extended secondary disk 
of very low surface brightness (\citealt{Thilker05, Barth07}).  The 
physical origin of these secondary outer-disk structures is not known, but a 
very intriguing possibility is that they might trace material accumulated from 
a recent episode of ``cold accretion.'' According to numerical simulations of 
cosmological structure formation (\citealt{Murali02}), nearly pristine, 
primordial gas should be constantly raining down on galactic disks as it 
condenses and cools from the hot halo, providing the raw material for 
continued disk building.

Lastly, non-axisymmetric perturbations in the disk play a primary role in
transferring angular momentum, thereby accelerating secular evolution by
redistributing the gas, and even the stars (e.g., \citealt{Friedli93, 
Athanassoula03}).  There are three main sources of
non-axisymmetric perturbations: bars and barlike oval structures, spiral arms,
and large-scale tidal distortions.  For example, the phenomenon of
``lopsidedness,'' which may be a manifestation of minor mergers, subtle tidal
interactions, or cold gas accretion (e.g., \citealt{Zaritsky97, Levine98,
Bournaud05}), would manifest itself as an $m=1$ Fourier
mode in azimuthal shape (e.g., \citealt{Peng10}).

The phenomena outlined above have all been investigated in the past to varying 
degrees.  Most previous studies have been quite restrictive, in terms of 
sample size, selection criteria, or wavelength coverage, often fine-tuned to 
address a narrow set of science goals.  Notable examples of early CCD-based 
surveys include those of \citet{Kormendy85}, \citet{Lauer85}, and 
\citet{Schombert86} on early-type galaxies, and those of \citet{Kent85}, 
\citet{Courteau96}, and \citet{deJong96} on spiral galaxies.  More diverse 
samples exist, but they are often 
targeted to probe specific environments (e.g., the field: \citealt{Jansen00}; 
nearby clusters: \citealt{Gavazzi03}).  Prior to the advent of modern all-sky 
surveys, two general-purpose samples of nearby galaxies have been widely 
used.  \citet{Frei96b} assembled optical images for 113 bright ($B_T$ \lax\ 
12.5 mag) galaxies spanning a wide range of Hubble types; although no rigorous 
selection criteria were applied, this catalog has been extensively adopted for 
a variety of studies because the authors removed foreground stars from the 
images \citep{Frei96} and released the digital atlas to the public.  The Ohio 
State University Bright Spiral Galaxy Survey (OSUBSGS; \citealt{Eskridge02}) 
enforced more rigorous selection criteria to define a set of 205 galaxies and 
expanded the photometric coverage to the near-infrared (NIR).  The lack of 
early-types in OSUBSGS compelled subsequent attempts to extend the sample to 
include S0 systems, but only in the NIR (\citealt{Laurikainen05, Buta06}).
Despite the utility of the Frei and OSUBSGS samples, they had 
limitations in terms of image quality and filter set.  Both surveys employed 
imager-telescope systems that resulted in rather coarse pixel scales, 
$\sim$1\farcs2-1\farcs4 pixel$^{-1}$ in the case of Frei, and 
$\sim$0\farcs4-0\farcs7 pixel$^{-1}$ in the case of OSUBSGS.  The image 
quality was especially heterogeneous for OSUBSGS, having been conducted using 
multiple detectors on six different telescopes.  

The considerations outlined above motivated us to initiate the 

\vskip 0.3cm
\psfig{file=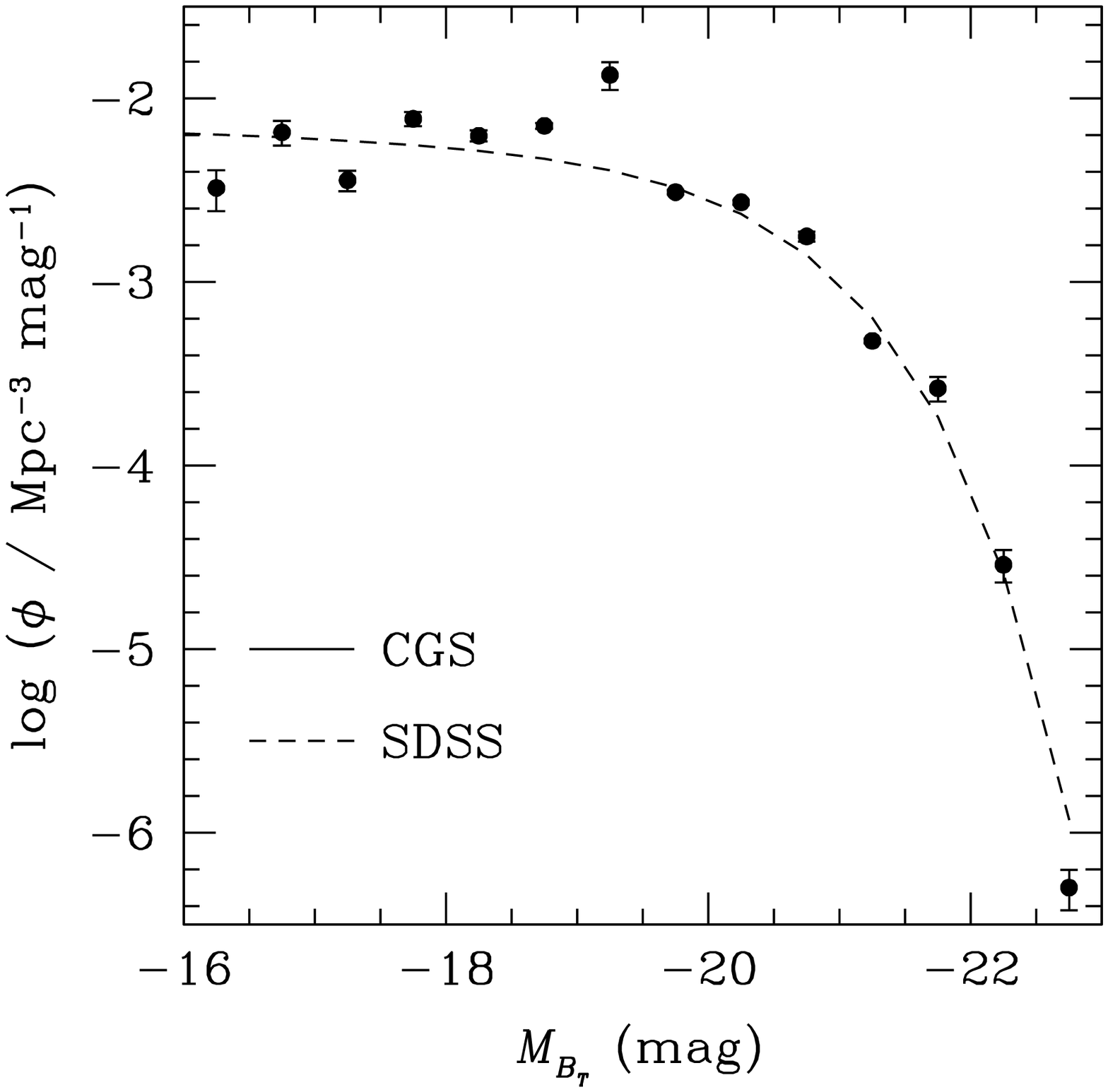,width=8.75cm,angle=0}
\figcaption[fig1.ps]{
$B$-band luminosity function for the CGS sample, computed with the
$1/V_{\rm max}$ method of \citet{Schmidt68}.  Superposed for comparison
is the $r$-band luminosity function of $z \approx 0.1$ galaxies selected
from SDSS (\citealt{Blanton03}), shifted by $B-r = 0.67$ mag, the average
color of an Sbc spiral (\citealt{Fukugita95}), which is roughly the
median Hubble type of the CGS sample.  The overall agreement indicates that
CGS provides an unbiased representation of the nearby galaxy population.
\label{fig1}}

\begin{figure*}[t]
\centerline{\psfig{file=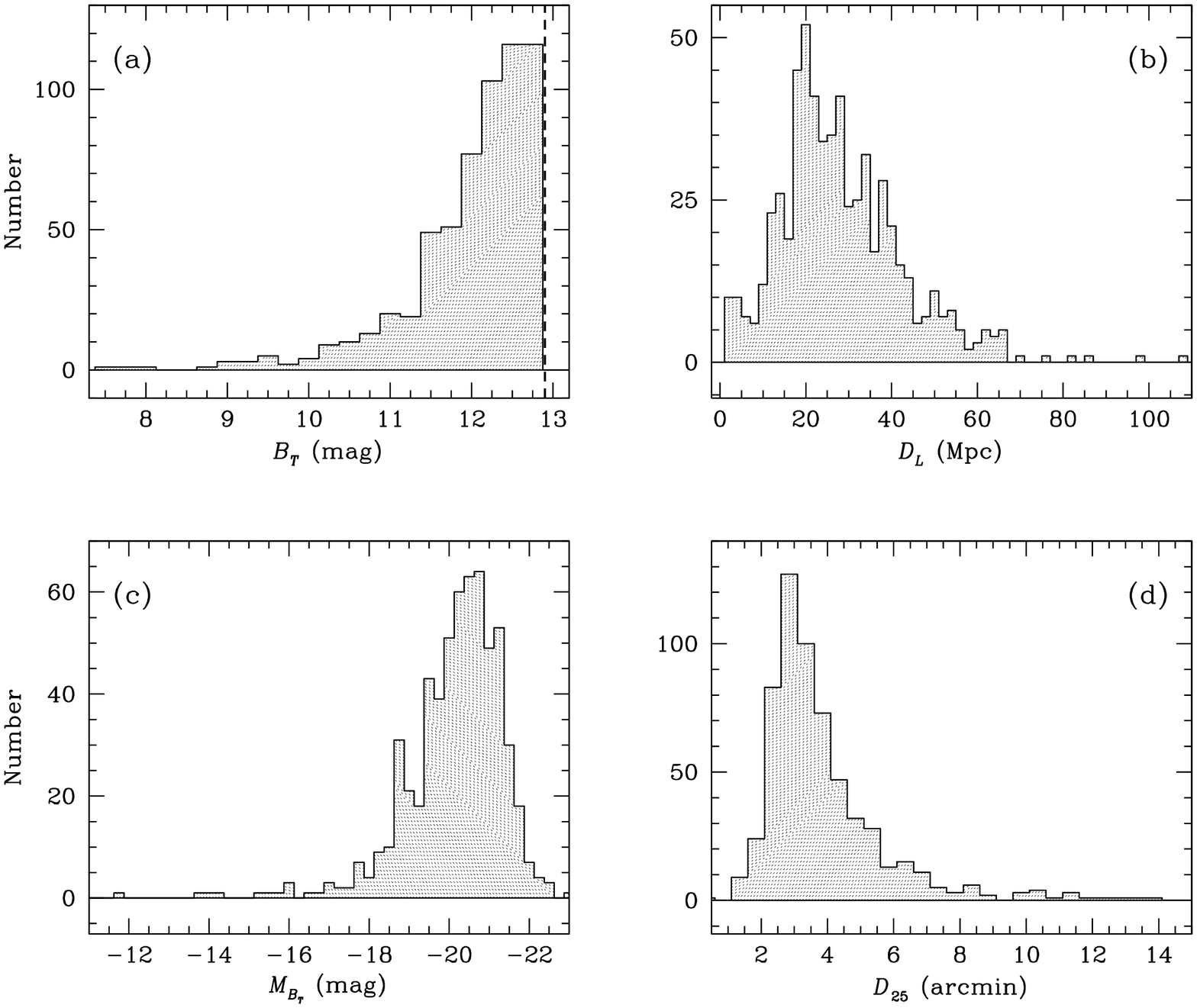,width=17.5cm,angle=0}}
\figcaption[fig2.ps]{
Basic properties of the CGS sample.  Distribution of (a) total
apparent $B$ magnitude, (b) distances, (c) absolute $B$ magnitude,
corrected for Galactic extinction, and (d) $B$-band isophotal diameter
at $\mu_B$ = 25 \sb.
\label{fig2}}
\end{figure*}

\noindent
Carnegie-Irvine
Galaxy Survey (CGS), a long-term program to investigate the
photometric and spectroscopic properties of a large sample of nearby galaxies,
using the facilities at Las Campanas Observatory.  Given the availability of
large databases such the Sloan Digital Sky Survey (SDSS; \citealt{York00}) or
the Two Micron All-Sky Survey (2MASS; \citealt{Skrutskie06}), it might seem
counterintuitive that a new galaxy imaging survey is necessary.  As described
below, our optical images have significantly better resolution than SDSS
images.  Moreover, only 9\% (56/605) of the galaxies in CGS overlap with the
SDSS footprint: SDSS does not cover the vast majority of bright southern
galaxies.  As the images are a vital precursor to the spectroscopic component
of the survey that we are planning to conduct, we have undertaken a uniform
imaging program for CGS.

This paper gives a general overview of the optical imaging component of CGS 
and summarizes a suite of ancillary data that will be used in subsequent 
analyses of the survey.  A companion paper by Li et al. (2011; hereafter 
Paper~II) describes the analysis of the surface brightness profiles and 
isophotal parameters of the sample.  Future papers will present the NIR imaging, 
detailed structural decompositions of the galaxies, and scientific 
applications thereof.

\section{Sample}

\subsection{Survey Definition}

To fully capture the wealth and range of the morphological properties of the
local galaxy population, to maximize signal-to-noise ratio (S/N) and spatial 
resolution, and to ensure statistical completeness, we target a large, 
well-defined sample of the brightest objects optimally placed for Carnegie's
facilities at Las Campanas Observatory.  We impose no 
selection according to galaxy morphology, size, or environment.  The sample, 
consisting of 605 objects, is formally defined by $B_T \leq 12.9$ mag and 
$\delta < 0$\deg.  The completeness level of bright galaxies to this magnitude 
limit is essentially 100\% (\citealt{Paturel03}), yielding a sample roughly 
comparable in size to that of the Palomar survey of northern galaxies 
(\citealt{Ho95, Ho97a}), which is desirable for future comparisons between the 
two hemispheres.  In practice, we select objects from the Third Reference 
Catalogue of Bright Galaxies (RC3; \citealt{deVaucouleurs91}), 

\begin{figure*}[t]
\centerline{\psfig{file=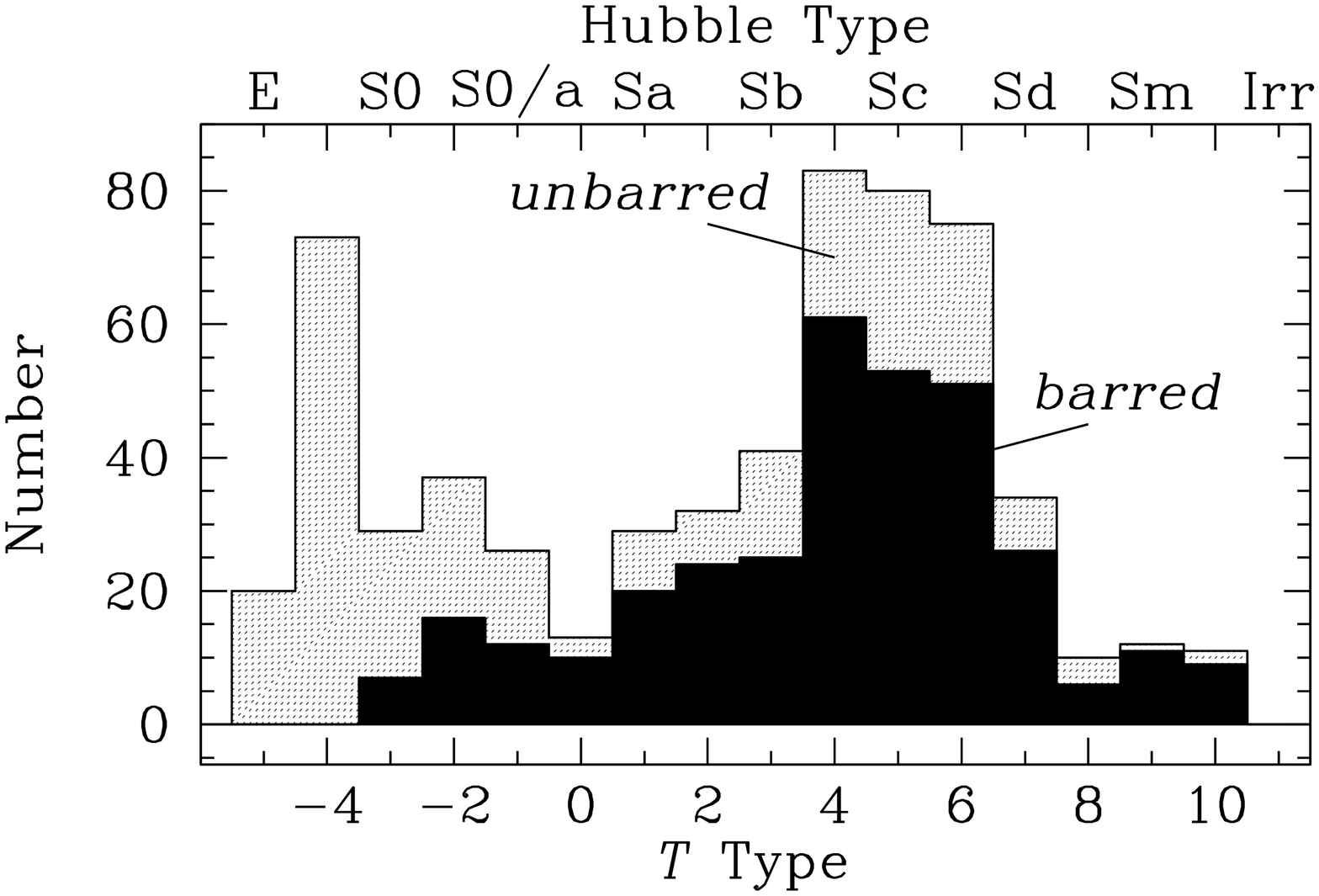,width=17.5cm,angle=270}}
\figcaption[fig3.ps]{
Distribution of morphological types in CGS.  The bottom axis gives the
type index $T$, and the corresponding Hubble types are shown on the top
axis (E: $T = -5$ to $-4$; S0: $-3$ to $-1$; S0/a: 0; Sa: 1; Sb: 3; Sc: 5; Sd:
 7; Sm: 9; Im: 10).  Barred (B and AB) and unbarred galaxies are shown in
filled and open histograms, respectively.
\label{fig3}}
\end{figure*}

\noindent
with the aid 
of the online database HyperLeda\footnote{{\tt http://leda.univ-lyon1.fr}} 
(\citealt{Paturel03}).  

CGS provides a fair, statistically unbiased representation of the local galaxy 
population over the absolute magnitude range $-16$ \lax\ $M_{B_T}$ \lax\ $-23$.
This is illustrated in Figure~1, which shows the $B$-band luminosity function 
calculated using the $1/V_{\rm max}$ method of \citet{Schmidt68}, compared with 
the $r$-band luminosity function of $z \approx 0.1$ galaxies derived from the 
SDSS by \citet{Blanton03}.  In making this comparison, we assume a typical
galaxy color for Sbc galaxies ($B-r = 0.67$ mag; \citealt{Fukugita95}), which 
is approximately the median morphological type of CGS, and a Hubble constant 
of $H_0 = 73$ \kms\ Mpc$^{-1}$.  The two functions agree quite well, both in 
shape and in normalization.  The relatively large fluctuations on the faint 
end of the luminosity function reflect the density inhomogeneities in the 
local volume.

Table~1 and Figure~2 summarize some of the basic parameters of the sample.  
Given the bright magnitude limit of the survey (Figure~2(a)), it is not 
surprising that most of the galaxies are quite nearby (median $D_L$ = 24.9 
Mpc; Figure~2(b)), luminous (median $M_{B_T} = -20.2$ mag, corrected for 
Galactic extinction; Figure~2(c)), and angularly large (median $B$-band 
isophotal diameter $D_{25}$ = 3\farcm3; Figure~2(d)).

The sample spans the full range of Hubble types in the nearby universe 
(Figure~3), comprising 17\% ellipticals, 18\% S0 and S0/a, 64\% spirals, and 
1\% irregulars.  A handful of well-known interacting systems (e.g., the 
``Antennae,'' Centaurus~A, NGC~2207) are included, but the vast majority of the 
sample have relatively undisturbed morphologies, although the RC3 
classifications designate $\sim$16\% of the sample as ``peculiar'' in some 
form.  As with the Palomar sample (\citealt{Ho97b}), roughly 2/3 of the 
disk (S0 and spiral) galaxies are barred according to the classifications in the 
RC3, with the breakdown into strongly barred (SB) and weakly barred (SAB) 
being 37\% and 29\%, respectively.  Approximately 10\% have an outer ring 
or pseudoring (\citealt{Buta91}).

In addition to the parent sample of 605 galaxies, Table~1 also lists 11 
sources that were observed early in the survey prior to it being fully 
defined, but that now formally do not meet the magnitude limit of the survey.
We include these extra objects for completeness, but they will be omitted from 
future statistical analyses.

\section{Observations}

All the images for CGS were taken with the du~Pont 2.5 m telescope, during 
the period 2003 February to 2006 June, spread over 69 nights and nine observing 
runs.  Table~2 gives a log of the observations.  For the optical component of 
the survey, we employed the 2048$\times$2048 Tek\#5 CCD camera, which has a 
scale of 0\farcs259 pixel$^{-1}$, sufficient to Nyquist sample the 
sub-arcsecond seeing often achieved with the du~Pont telescope.  The 
field of view of 8\farcm9$\times$8\farcm9 is large enough to enclose most of 
the galaxies, which have a median isophotal diameter of $D_{25}$ = 3\farcm3 at 
a surface brightness level of $\mu_B=25$ mag arcsec$^{-2}$ (Figure~2(d)), while 
still allowing adequate room for sky measurement.  As discussed in Section~4.4, 
background subtraction is a key factor that limits the accuracy of structural 
decomposition and detection of faint outer features.  A subset of the larger
galaxies were reimaged at lower resolution with the Wide-field CCD, which
has a field of view of 26\amin$\times$26\amin\ and a scale of 
0\farcs77~pixel$^{-1}$; these data will be presented elsewhere.  

Each galaxy was imaged in the Johnson $B$ and $V$ and Kron--Cousins $R$ and $I$ 
filters, typically for total integration times of 12, 6, 4, and 6 minutes, 
respectively, split into two equal-length exposures to facilitate rejection 
of cosmic rays and to mitigate saturation.  The centers of some galaxies were 
still saturated even with these integration times, and we took short ($\sim 
10-60$ s) exposures to obtain unsaturated images of the core.  In total, 
over 6000 science images were collected.  Standard calibration frames for 
optical CCD imaging were taken, including a series of bias frames, dark 
frames, and flat fields from the illuminated dome and the twilight sky.  
During clear nights, we observed a number of photometric standard star fields 
from \citet{Landol92}, covering stars with a range of colors and tracking over a 
range of airmasses.

A significant fraction of the sample was also imaged in the $K_s$ (2.2 
\micron) band using the Wide-field Infrared Camera, a cryogenically 
cooled mosaic of four 1024$\times$1024 arrays that delivers 
13\amin$\times$13\amin\ images with excellent image quality (scale 0\farcs2 
pixel$^{-1}$). These observations will be presented elsewhere.

\section{Data Reductions}

\subsection{Basic Processing}

The initial processing follows standard steps for CCD data reduction within 
the IRAF\footnote{IRAF is distributed by the National Optical Astronomy 
Observatory, which is operated by the Association of Universities for Research 
in Astronomy (AURA), Inc., under cooperative agreement with the National Science 
Foundation.} environment, using tasks within {\it ccdproc}, which
include trimming, overscan correction, bias subtraction, and flat fielding.  
Dark current subtraction is unnecessary.  We took dark frames with integration 
times comparable to those of the longest science exposures, but the dark 
current always turns out to be negligible.  For each night of each observing 
run, we generate a master bias frame by averaging a large number (typically 
$\sim 20$) of individual bias frames.  Similarly, we create a master flat-field 
frame with high S/N by combining a series (typically 
$\sim 6-10$ per filter) of domeflats and twiflats.  Through experimentation, 
we found that the twiflats produce more uniform illumination than the 
domeflats for the $B$ and $V$ images, whereas the domeflats are more effective 
for the $R$ and $I$ images.  A small fraction of the flats contained dust 
specks that introduced ``doughnut'' features into the flattened frames; 
roughly $\sim 4$\% of the science images were affected by this.  The flattened 
$I$-band images contain subtle, residual fringe patterns with amplitudes (peak 
to trough) at the level of $\sim$1\%--2\%.  We remove these by subtracting an 
optimally scaled fringe frame with a mean zero background, which was 
constructed by median combining a total of 36 sky images collected over many 
observing runs.  The fringe pattern proved to be remarkably stable throughout 
the course of the survey.  Apart from a couple of partially dead columns near 
the center of the chip, the Tek\#5 CCD is relatively clean cosmetically.  We 
generate a mask for these and other less conspicuous defective regions, and 
use it for local interpolation to correct the bad pixels.   Finally, we use an 
IDL version of van~Dokkum's (2001) L.A.Cosmic routine to correct the pixels 
affected by cosmic ray hits and satellite trails; any remaining imperfections 
were further edited manually.  

Next, we shift and align the flattened, cleaned images in each filter 
to a common reference frame defined by the $I$-band image, which generally 
has the sharpest seeing and is least affected by possible dust obscuration.  
Multiple images taken with the same filter were combined.  We carefully 
examine the images for possible saturation near the core of the galaxy and 
replace any saturated pixels with their appropriately scaled counterparts from 
the short-exposure images.  We determine the proper scale factor by comparing
the background-subtracted, integrated counts of bright, isolated field stars 
that are unsaturated in both the long and short exposures. The affected 
regions are generally small, spanning a diameter of only $\sim 10$ pixels.  As 
this is not significantly larger than the point-spread function (PSF), this 
procedure has a negligible impact on any of our subsequent scientific 
analysis.  For convenience, we flip the images so that north points up and 
east to the left.  Lastly, we use the \emph{Astrometry.net} (Lang et al. 2010) 
software\footnote{{\tt http://www.astrometry.net}} to solve for the World 
Coordinate System (WCS) coordinates of each image and store them in the FITS 
header.

\subsection{Photometric Calibration}

A little more than half of the CGS galaxies were observed under 
photometric conditions, as determined from observations of \citet{Landolt92} 
standard stars.  We measure the photometry of the stars using a circular 
aperture with a fixed radius of 7\asec, to mimic as closely as possible 
Landolt's measurements.  We determine the local sky of each star from a 
5~pixel wide annulus outside that aperture, using only sky pixels (i.e.,
avoiding any nearby neighboring stars).  The median photometric errors 
(Table~2, Column 9) for the photometric nights are 0.08, 0.04, 0.03, and 0.04 
mag for the $B$, $V$, $R$, and $I$ band, respectively.  The images calibrated 
in this manner have the header keyword {\tt ZPT\_LAN}.

For the non-photometric observations, we adopt an indirect approach that 
allows us to obtain a less accurate, but still useful, photometric
calibration.  Our strategy is to bootstrap the instrumental magnitudes of the 
brighter stars within each CCD field to their 
corresponding photographic magnitudes as published in the second-generation 
\emph{Hubble Space Telescope} Guide Star Catalog (GSC2.3; \citealt{Lasker08}).  
We derive the transformation between the GSC photographic bandpasses and 
our standard \emph{BVRI} magnitudes using the subset of field stars that were 
observed by us under photometric conditions.  The GSC contains astrometry, 
photometry, and object classification for nearly a million objects.  The 
photometry for the southern hemisphere is given primarily in three 
photographic bandpasses, $B_J$, $R_F$, and $I_N$. The typical photometric 
error for the stellar objects in the GSC, depending on the magnitude and 
passband, is 0.13--0.22 mag (\citealt{Lasker08}).

Starting with the list of WCS coordinates extracted for the field stars in 
each CCD image, we use the Vizier server\footnote{{\tt 
http://vizier.cfa.harvard.edu/viz-bin/VizieR}} to query the GSC for objects 
classified therein as stellar ({\tt class = 0}).  We eliminate 
saturated objects and stars fainter than $B \approx 22$ mag.  To 
properly compare our instrumental magnitudes with the GSC magnitudes, it is 
critical that we follow as closely as possible the methodology that the GSC 
used to derive their photometry.  We cannot use conventional methods of 
stellar aperture photometry (e.g., using standard tasks in IRAF).  As outlined 
in \citet{Lasker08}, the procedures used in the GSC for object detection, 
flux measurement, sky determination, and source deblending in the case of 
crowded fields (\citealt{Beard90}) follow very closely those 
implemented in the SExtractor package (\citealt{Bertin96}).  We 
therefore use SExtractor to measure the stars in our CGS images.  We manually 
performed aperture photometry of a number of isolated stars to confirm that 
the SExtractor-based measurements are reliable, to better than 0.02 mag.

We intercalibrate the CGS and GSC photometric scales using data from a night 
with exceptional photometric stability, during which the open cluster M67 was 
observed.  Our photometry  for M67 agrees very well with that published by 
\citet{Montgomery93}, to better than 0.01 mag, for all filters.  Using a 
set of $500-600$ field stars selected from the science images observed 
throughout the course of that night, we derive a set of transformation 
equations between the GSC photometric bandpasses ($B_J$, $R_F$, and $I_N$) and 
our standard bandpasses ($B$, $V$, $R$, and $I$).  The fitting was done with 
the IDL function \emph{ladfit}, which uses a robust least absolute deviation 
method and is not sensitive to outliers.  The transformation equations and 
residual scatter ($\sigma$) (in magnitudes) for each filter are as follows:

\begin{eqnarray}
B &=& B_J + 0.2807 - 0.1365 (B_J - R_F) \quad \qquad  \sigma = 0.2088 \\
B &=& B_J + 0.1499 + 3.239 \times 10^{-5} (B_J - I_N) \quad \sigma = 0.2490 \\
B &=& 0.1489 + 1.0001 B_J  \ \ \ \ \ \ \qquad\qquad\qquad \sigma = 0.2597 \\
B &=& R_F + 0.9700 + 0.8888 (R_F - I_N)   \ \ \ \ \qquad \sigma = 0.2982 \\
B &=& 0.4508 + 1.0468 R_F   \ \ \ \ \ \qquad\qquad\qquad \sigma = 0.3832 \\
B &=& 0.8471 + 1.0488 I_N   \ \ \ \ \ \ \qquad\qquad\qquad \sigma = 0.6058
\end{eqnarray}

\begin{eqnarray}
V &=& B_J + 0.1512 - 0.6625 (B_J - R_F)  \qquad\quad \sigma = 0.1140 \\
V &=& R_F + 0.4658 + 0.2279 (R_F - I_N)  \qquad\quad \sigma = 0.1721 \\
V &=& B_J + 0.0060 - 0.3982 (B_J - I_N)  \ \qquad\quad \sigma = 0.1760 \\
V &=& 0.8077 + 0.9825 R_F  ~ \ \ \ \ \ \qquad\qquad\qquad \sigma = 0.1828 \\
V &=& 0.7665 + 0.9232 B_J  \ \ \ \ \ \ \qquad\qquad\qquad \sigma = 0.2775 \\
V &=& 0.0875 + 1.0499 I_N  ~ \ \ \ \ \ \ \qquad\qquad\qquad \sigma = 0.3780
\end{eqnarray}

\begin{eqnarray}
R &=& R_F + 0.1380 - 0.2667 (R_F - I_N)  \qquad\quad\sigma = 0.0976 \\
R &=& 1.2166 + 0.9286 R_F  \ \ \ \ \ \ \qquad\qquad\qquad\sigma = 0.1121 \\
R &=& I_N - 0.0954 + 0.3353 (B_J - I_N) ~ \ \qquad\quad\sigma = 0.1257 \\
R &=& R_F - 0.0208 + 0.0551 (B_J - R_F)  \qquad\quad\sigma = 0.1264 \\
R &=& -0.2960 + 1.0441 I_N  ~ ~ \ \ \ \ \qquad\qquad\qquad \sigma = 0.2301 \\
R &=& 1.5288 + 0.8546 B_J   \ \ \ \ \ \ \ \qquad\qquad\qquad\sigma = 0.3509
\end{eqnarray}

\begin{eqnarray}
I &=& I_N - 0.1119 + 0.0931 (R_F - I_N) \ \qquad\quad\sigma = 0.0924 \\
I &=& I_N - 0.0709 + 0.0152 (B_J - I_N) ~ \ \qquad\quad\sigma = 0.1073 \\
I &=& -0.1713 + 1.0079 I_N  \ \ \ \ \  \qquad\qquad\qquad\sigma = 0.1164 \\
I &=& R_F - 0.2323 - 0.2276 (B_J - R_F) \qquad\quad\sigma = 0.2000 \\
I &=& 1.6690 + 0.8728 R_F \ \ \ \ \ \ \qquad\qquad\qquad\sigma = 0.2244 \\
I &=& 2.5445 + 0.7739 B_J ~ ~ \ \ \ \ \ \qquad\qquad\qquad\sigma = 0.4543
\end{eqnarray}

\bigskip
For any given CCD frame, the calibration equation we choose depends on which 
GSC bandpasses are available for that particular field; all else being equal, 
we select the transformation equation that has the smallest scatter.  The 
images calibrated in this manner have the keyword {\tt ZPT\_GSC} in 
the header.  The median photometric uncertainty (Table~2, Column 9) of the 
GSC-based magnitudes, which is dominated by the error in the above fitting 
equations, is 0.21, 0.11, 0.098, and 0.092 mag for the $B$, $V$, $R$, and $I$ 
band, respectively.

\subsection{Masks and Point-spread Function}

Many aspects of the survey require that we exclude signal contributed by 
unrelated objects, such as foreground stars and background galaxies.  This is 
achieved by creating a mask that properly identifies the affected pixels.  We 
begin by running SExtractor to identify all the objects in the image and
classify them into stellar and nonstellar sources based on their compactness 
relative to the seeing.  Through experimentation, we find that a ``class 
parameter'' larger than 0.3 effectively isolates the stars from the resolved 
objects.  During this initial step, we deliberately set the contrast parameter 
to a moderate value (we set the threshold to be 3 $\sigma$ above the 
background) so that galaxies with substantial internal structure (e.g., 
spirals and irregulars) do not get broken up into multiple pieces.  This, in 
turn, implies that SExtractor will miss fainter stars that might be superposed 
on the main body of the central galaxy.  We carefully examine each image 
visually and manually add additional objects to the segmentation image if 
necessary.  Next, two segmentation images are created, one to isolate the 
unsaturated stellar objects and the other the nonstellar objects, which 
include background galaxies and saturated stars (including their bleed trails, 
if present).  Because we set the contrast parameter to a conservative level, 
the segmentation images identify only the bright core regions of the objects 
to be masked and often miss their fainter outer halos, which can be 
substantial, especially for bright stars in the $I$ band.  To account for 
these regions, we need to ``grow'' the segmentation image of each object.  
From trial and error, we find that a growth of 8 pixels in radius is optimal 
for the stellar objects.  For the nonstellar objects, we distinguish between 
two regimes.  A growth radius of 5 pixels is sufficient for bright bleed 
trails.  For the fainter, more extended halos, we find that the growth radius 
can be approximated by $R={\rm min}[15, 1.5 \ \sqrt{N/\pi}]$, where $N$ is the 
number of pixels contained in the original segmentation image.  Lastly, we 
stack together the masks of each individual filter to create a master mask for 
each galaxy, which is then used in all subsequent analysis, to ensure that all 
filters reference the same set of pixels.

We use the IRAF task \emph{psf} within the \emph{daophot} package and the 
bright, unsaturated stars identified by SExtractor in the previous step to 
build an empirical PSF for each image.  The PSF image is a key ingredient in 
much of our analysis, including generation of the structure maps (Section~5) 
and bulge-to-disk decomposition (S. Huang et al., in preparation).

\subsection{Image Quality}

The bulk of the CGS observations have fairly good image quality.   The seeing, 
as estimated from the full width at half-maximum (FWHM) of the radial profiles 
of bright, unsaturated stars, ranges between $\sim$0\farcs5, the limit we can 
measure given 

\vskip 0.3cm
\psfig{file=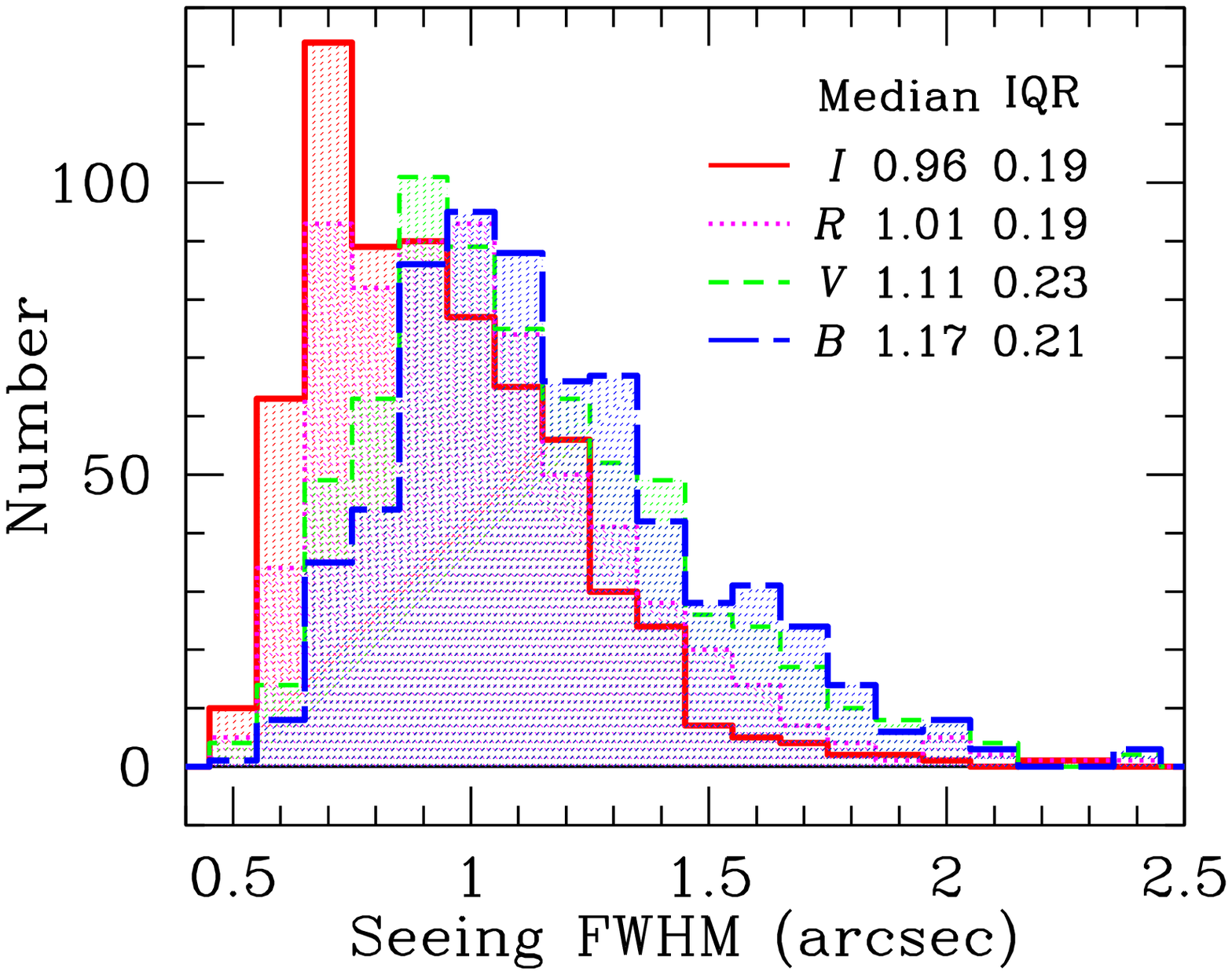,width=8.75cm,angle=0}
\figcaption[fig4.ps]{
Distribution of seeing values for the CGS images, shown separately
for each of the four filters.  The legend gives the median and interquartile
(IQR) range of each histogram.
\label{fig4}}
\vskip 0.3cm

\noindent
the pixel scale of our detector, to $\sim 2$\asec, with a median 
value of 1\farcs17, 1\farcs11, 1\farcs01, and 0\farcs96 in the $B$, $V$, $R$, 
and $I$ band, respectively (Figure~4). The main factors that limit the surface 
brightness sensitivity of the images stem from residual errors in 
flat fielding and uncertainties in sky determination.  We have discovered 
that large variations in the ambient temperature and humidity throughout the 
night can occasionally produce water condensation on the surfaces of some of 
the filters.  This effect imprints large-scale flat-fielding errors at the 
level of $\sim 1\%$.   Background gradients of a roughly similar magnitude are 
sometimes induced by scattered light from very bright stars, either within or 
just outside the science frame.  These anomalies can be mitigated by fitting a 
two-dimensional polynomial function to the background pixels 
(\citealt{Noordermeer07, Barazza08, Erwin08}).  A second-order 
polynomial function usually suffices, but in some cases the order of the 
polynomial may need to be as high as 5.  The fitting regions are source-free 
pixels far from the main galaxy, selected from the image after convolving it 
with a Gaussian with FWHM $\approx$ 4\asec--5\asec\ to accentuate low-spatial 
frequency features.  This correction typically improves the flatness of the 
images by about a factor of 2, from $\sim 1\%$ to $\sim 0.6\%$.  Once the 
background is flat, determining its value is relatively straightforward 
because most of the survey galaxies have sizes (median $D_{25}$ = 3\farcm3; 
Figure~2(d)) that fit comfortably within the field of view of the detector 
(8\farcm9$\times$8\farcm9).  The larger galaxies ($D_{25}$ \gax\ 
5\amin--6\amin; $\sim$15\% of the sample), however, pose a challenge, and we 
are forced to implement an indirect, less reliable strategy for sky 
subtraction (Paper~II).  The surface brightness depth of the images (Table~2, 
Column 10), which we define to be the isophotal intensity that is 1 $\sigma$ 
above the sky rms, has a median value of $\mu \approx 27.5$, 26.9, 26.4, and 
25.3 mag~arcsec$^{-2}$ in the $B$, $V$, $R$, and $I$ bands, respectively.

\section{Images}

In anticipation of future applications of the survey, we prepare 
several different renditions of the digital images, which are described 
below.  The full image atlas for the 605 galaxies in CGS (including the 11 
``extras'') is given in the Appendix, as Figures~7.1--7.616, as well as on the 
project Web site {\tt http://cgs.obs.carnegiescience.edu}.

\begin{itemize}

\item{{\it Color composites}.  We generate three-color composites from the 
$B$, $V$, and $I$ images by applying an arcsinh stretch (\citealt{Lupton04}) 
to each of these bands and then using them to populate the blue, green, and 
red channels of the color image.  The arcsinh stretch is particularly 
effective in viewing a broad dynamic range of structure.
}

\item{{\it Star-cleaned images}.  A variety of scientific applications can 
benefit from galaxy images that are free from contamination by foreground 
stars and background galaxies.  An example would be using nearby galaxies as 
templates to simulate higher redshift observations.  Our procedure to 
``clean'' the galaxy images begins by using the object mask created from 
SExtractor (Section~4.3) to identify all the stellar and nonstellar objects
that need to be removed.  For every given image, we use the task {\em allstar}
to fit the empirical PSF appropriate for that image to all the unsaturated 
stars and subtract them.  As the PSF varies slightly across the field, the 
subtraction is imperfect and small residuals often remain around the 
positions of bright stars.  To improve the cosmetic appearance of the images, 
we devised the following scheme to locally interpolate over the residuals. 
For every star, we fit a two-dimensional function to a 5 pixel wide annular 
region immediately exterior to its grown segmentation image.   We find the 
exponential function to be quite effective---more so than a tilted plane or a 
second-order polynomial---as it has enough flexibility to handle steep local 
gradients often encountered for complex galaxy backgrounds.  Each pixel within
the segmentation image is then compared with the best-fitting function value 
at that position.  If they differ by more than 2 times the mean value of the 
differences between the pixels in the background annulus and the best-fit 
pixel values there, it gets replaced with the best-fit pixel value.  We add 
noise to the replaced pixel, drawn randomly, using the IDL function 
{\em randomu}, from the distribution of residuals calculated from the 
background annulus.  The nonstellar objects are treated in exactly the 
same manner, except that for these PSF subtraction is unnecessary.

The above procedure works very effectively most of the time, as illustrated
for a couple of examples in the top and middle panels of Figure~5.  It 
performs less successfully for heavily saturated bright stars that are 
directly superposed on the science target, especially those with prominent 
bleed trails, as shown in the example in the bottom panel of the figure.
Under these circumstances, almost any local interpolation scheme will produce 
a highly compromised result.  For these situations, we extract, in 
conjunction with the SExtractor object mask,  the surface brightness profile 
and isophotal parameters of the galaxy using the IRAF task {\it ellipse}, as 
described in Paper~II.  These isophotes then serve as the basis for building 
a smooth representation of the intrinsic light distribution of the galaxy 
using the task {\it bmodel}, which, when combined with the original object 
mask, gives a reasonably realistic estimate of the underlying galaxy light 
affected by the masked regions.  The corrupted regions were replaced with
the corresponding pixel values from the model image, and Poisson noise is 
added.  If a saturated star falls near the center of the galaxy, we 

\begin{figure*}[t]
\centerline{\psfig{file=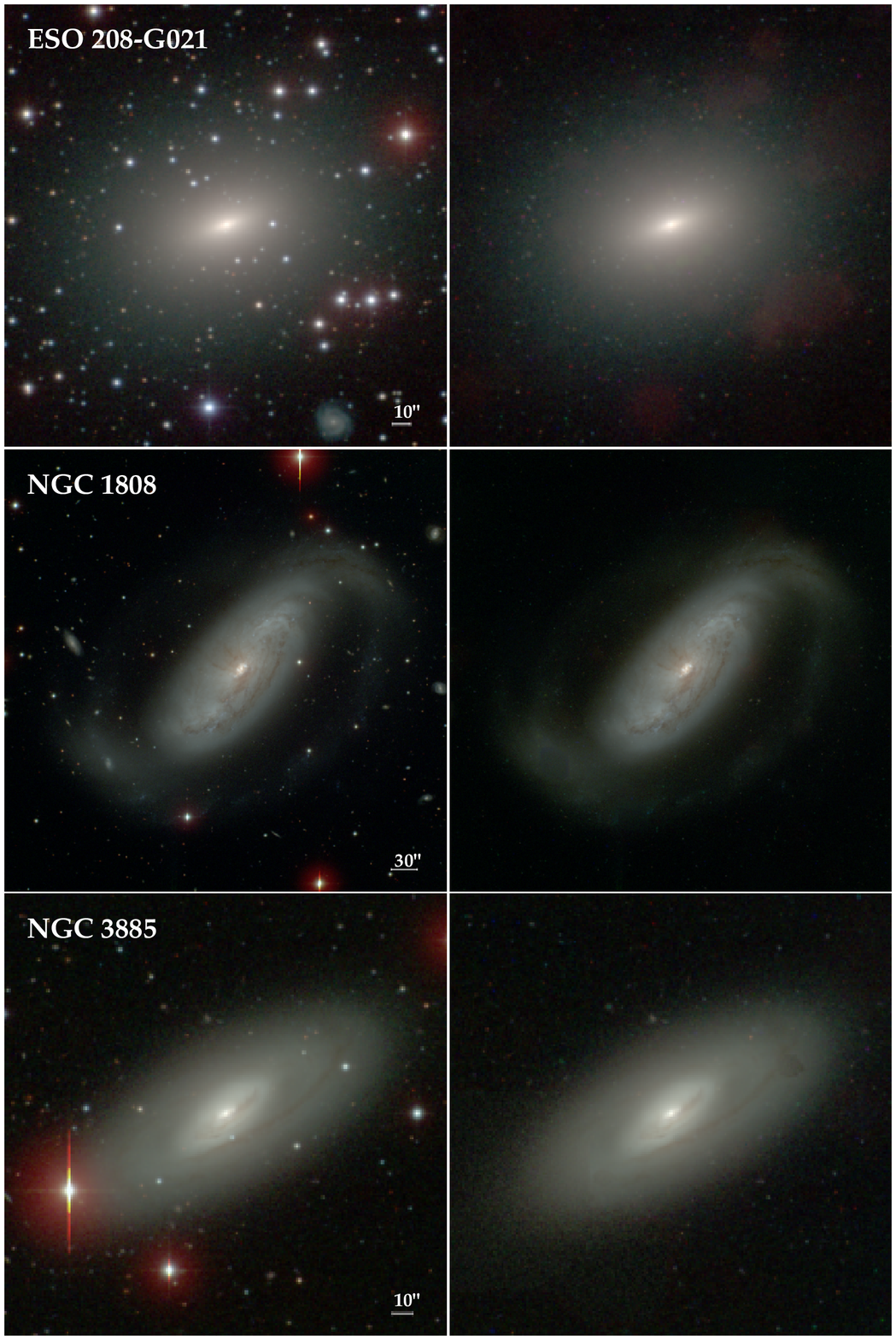,width=17.0cm,angle=0}}
\figcaption[fig5.ps]{
Examples to illustrate our technique to clean the images of
foreground stars and background galaxies.  We use the three-band composite
for both the original and star-cleaned images, displayed using an arcsinh
stretch.
\label{fig5}}
\end{figure*}

\begin{figure*}[t]
\centerline{\psfig{file=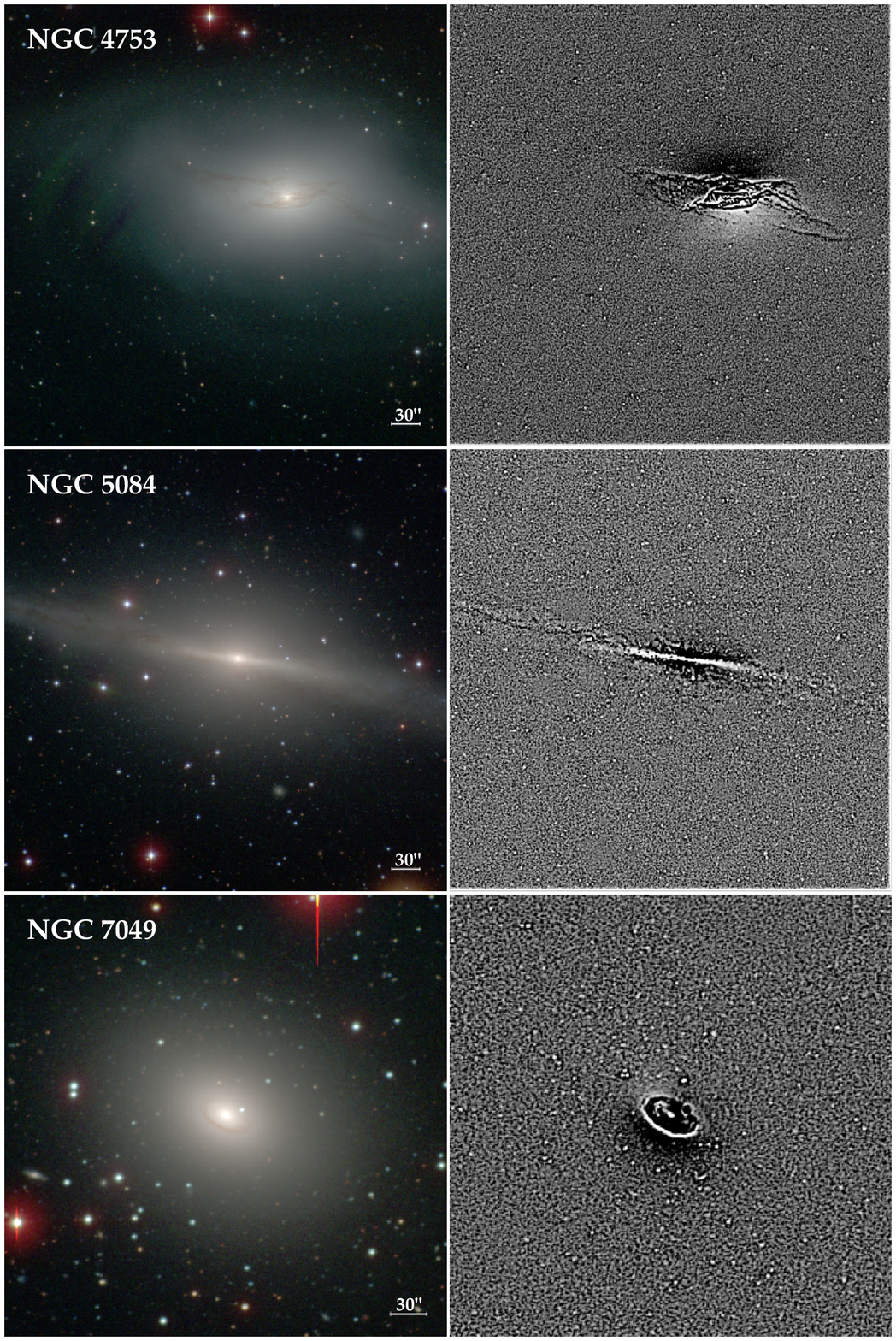,width=17.0cm,angle=0}}
\figcaption[fig6.ps]{
Examples to illustrate the ability of the structure maps to bring out
fine structure.  The left panel shows the \emph{BVI} color composite on an
arcsinh stretch, and the right panel shows the structure map of the
star-cleaned $B$-band image on a linear stretch.
\label{fig6}}
\end{figure*}

\clearpage

\noindent
first
attempt to replace it with an unsaturated version using the short exposure,
following the same procedure for treating saturated cores.  We apply this
alternative procedure only to those regions wherein our local interpolation
scheme fails to do a good job.
}

\item{{\it Stacked images}.  In an effort to improve the S/N and to enhance 
the visibility of features with low surface brightness, we create a stacked 
image for each galaxy by combining the processed individual images from each 
of the four filters.  We did not match the seeing among the images, as such 
a process, which involves smoothing, would alter the noise properties of the 
images; this is not critical for the stacked images because we are mainly 
interested in emphasizing diffuse features, which are insensitive to the 
seeing.  This technique has been used to detect faint structures in both 
nearby (\citealt{vanDokkum05}) and distant (\citealt{Gawiser06}) galaxies.  As in 
\citet{Gawiser006}, their Appendix~A), we weight the data by S/N$^2$ to 
optimize for surface brightness, after first normalizing all the images so 
that 1 count corresponds to AB = 30 mag.   We follow \citet{Frei94} to 
convert our \emph{BVRI} magnitudes to the AB system.  To compute the weights, 
we determine the signal from the mean flux of 10--20 bright stars in common 
among the different images, and the noise from the standard deviation of the 
sky pixels.
}

\item{{\it Structure maps}.  There are a variety of ways to spatially filter 
an image to emphasize structures on different scales.  The structure map 
technique developed by \citet{Pogge02} is particularly effective in 
enhancing spatial variations on the smallest resolvable scales, namely that 
of the PSF.  For every image in the survey, we calculate the structure map 
$S$, defined as 

\begin{equation}
S = \left({{I}\over{I \otimes P}}\right)\otimes P^{\phantom{.}t},
\end{equation}

\noindent
where $I$ is the star-cleaned image taken in a particular filter, $P$ is the 
corresponding PSF image (see Section~4.3), $P^{\phantom{.}t}$ is the transpose 
of the PSF, such that $P^{\phantom{.}t}(x,y) = P(-x,-y)$, and the operator 
$\otimes$ denotes convolution.  Figure~6 illustrates the utility of the 
structure map in highlighting small-scale spatial variations in an image.
}

\item{{\it Color index maps}.  Our multi-filter data set allows us to study
two-dimensional distributions of color, which trace spatial variations of dust 
and stellar content.  A color index map is generated simply by dividing two 
registered images taken in the relevant filters, after sky subtraction and 
matching the two images to a common PSF.  We convolve the image with the 
better seeing with a two-dimensional Gaussian function whose width is the 
quadrature difference of the two seeing values.  We create the following set 
of color index maps: $B-V$, $B-R$, $B-I$, $V-I$, and $R-I$.
}
\end{itemize}

\section{Tabulated Data}

\subsection{Isophotal and Photometric Parameters Derived from CGS}

Paper~II presents the brightness profiles and isophotal parameters of the 
survey.  A number of basic, but nonetheless useful, global parameters for the 
galaxies can be derived from those data.  We summarize them here.  The 
advantage of our measurements is that they are derived in a uniform, 
self-consistent manner. 

Table~3 lists several size measurements derived from the $B$-band images.  We 
choose this bandpass because it closely matches most published measurements 
in the literature.  The quantities $R_{20}$, $R_{50}$, and $R_{80}$ are the 
radii enclosing, respectively,  20\%, 50\%, and 80\% of the total flux, which 
is calculated from integrating the surface brightness profile generated using 
the IRAF task \emph{ellipse} (see Paper~II).  We define the concentration 
parameter as $C = R_{80}/R_{20}$.  The isophotal diameters, measured at a 
surface brightness level of $\mu_B = 25.0$ and 26.5 mag~arcsec$^{-2}$, are 
given as $D_{25}$ and $D_{26.5}$, while $D^c_{25}$ and $D^c_{26.5}$ are the 
corresponding values corrected for Galactic extinction ($A_B$ in Table~1, 
taken from \citet{Schlegel98} and inclination effect, following the 
prescription of \citet{Bottinelli95}.
For galaxies with morphological types $T > -3.5$, we 
derive the inclination angle of the disk, $i$, using Hubble's (1926) 
formula

\begin{equation}
{\rm cos}^2  i = {{(1-e)^2 - q_0^2}\over{1-q_0^2}},
\end{equation}

\noindent
which makes use of the apparent ellipticity ($e$) of the outer regions of the 
disk. The intrinsic flattening of the disk depends mildly on morphological type 
(\citealt{Paturel97}), but for simplicity we adopt a constant value of $q_0 = 
0.2$ (\citealt{Noordermeer07}).  The position angle (PA) of the 
photometric major axis, east of north, is also given.  Unlike the other 
parameters in this table, both $e$ and PA are measured from the $I$-band image 
rather than from the $B$-band image in order to minimize possible distortions 
from dust or patchy regions of star formation.  These quantities are 
determined from the outer regions of the galaxy---far enough to be insensitive 
to the bulge and bar but not so much so to be affected by lopsidedness in the 
outer disk or imperfect flat fielding---where they usually converge to a 
constant value.  The listed uncertainties represent the standard deviation 
about the mean.  In cases where $e$ and PA do not converge, we estimate them 
visually from the best-fitting isophotes at large radii.  Lastly, we tabulate 
$\Sigma$, the sum of the rms fluctuations in the structure map of the $B$-band 
image, which gives a useful relative measure of the degree of high-spatial 
frequency fluctuations present in the image.  

Table~4 presents \emph{BVRI} integrated magnitudes derived from the 
star-cleaned images (Section~5). Two measurements are made: $m_{25}$ pertains 
to the total apparent magnitude within the $\mu = 25$ mag~arcsec$^{-2}$ 
isophote, whereas $m_{\rm tot}$ derives from the total flux within the largest 
isophote fitted to the image.  The main uncertainty of the magnitudes comes 
from the uncertainty of the photometry zero point, with little contribution 
from sky error or Poisson noise.  For convenience, we provide the absolute 
magnitudes in the $B$ band, corrected for Galactic extinction as given in 
Table~1 and adopting the luminosity distances therein.

Table~5 lists $\mu_e$, the surface brightness measured at the effective 
or half-light radius $R_{50}$, for each of the four filters.

\subsection{Photometric Data from the Literature}

Table~6 assembles multiwavelength photometric data from several large 
galaxy catalogs.  The listed values represent total integrated fluxes for 
the entire galaxy.  Instead of collecting data from any available literature 
source, we restrict our selection to only a few well-documented, homogeneous
compilations.  For the ultraviolet, we select FUV (1350--1750 \AA) and NUV 
(1750--2800 \AA) measurements principally from the {\it Galaxy Evolution 
Explorer}\ ({\it GALEX}) atlases of 
\citet{Buat07}, \citet{Dale07}, and \citet{GildePaz07}.  At 
present, only $\sim 16\%$ of our sample has published {\it GALEX}\ data, 
although the situation will soon improve substantially with the recent 
completion of the All-Sky Imaging Survey\footnote{\tt 
http://galex.stsci.edu/GR6}.  To extend the optical coverage blueward of the 
$B$ band, we searched HyperLeda for $U$-band (3500 \AA) photometry, which is 
available for 63\% of the sample.  The most complete coverage is in the 
near-infrared ($J$, $H$, and $K_s$ bands) and far-infrared (12, 25, 60, and 100 
\micron); $\sim$95\% of the galaxies are included in the Two Micron All-Sky 
Survey (2MASS; \citealt{Skrutskie06}) and in the source catalogs of the 
{\it Infrared Astronomical Satellite}\ ({\it IRAS}).  Because our 
galaxies are angularly large, care is taken to choose photometry derived from 
studies that properly treat extended sources (\citealt{Jarrett03, Sanders03}).
In addition to the flux densities in the individual {\it IRAS}\ 
bands, we also give the parameter FIR, defined as 
1.26\e{-14}\,(2.58$S_{60}\,+\,S_{100}$) W~m$^{-2}$, which approximates well the 
total flux between 42.5 and 122.5 \micron\ (\citealt{Helou88, Rice88}).
To date, $\sim 70$\% (406/605) of our sample have been observed at 3.6 
and 4.5 \micron\ using the Infrared Array Camera on {\it Spitzer}.  Roughly 
60\% of our sample overlaps with the {\it Spitzer}\ Survey of Stellar 
Structure in Galaxies (S$^4$G; \citealt{Sheth10}).  We intend to incorporate 
these mid-infrared data into CGS in the future.

\subsection{Kinematics, Environment, and Gas Content}

Three categories of data are included in Table~7.  We assemble from HyperLeda
information on the internal kinematics of the galaxies, namely the apparent 
maximum rotation velocity of the gas ($V_{\rm max}$), the maximum rotation 
velocity corrected for inclination ($V_{\rm rot}$), and the central stellar 
velocity dispersion ($\sigma_{\star}$).  

Next, we provide three environmental 
indicators.  Using the facilities in the NASA/IPAC Extragalactic Database
(NED)\footnote{\tt http://nedwww.ipac.caltech.edu}, we search for candidate 
neighbors within a radius of 750 kpc.  (NED restricts environmental searches 
to a maximum radius of 600\amin, and, for a handful of objects, this 
corresponds to less than 750 kpc.) We calculate the projected angular 
separation, $\Delta \theta$, in units of the diameter $D_{25}$ (Table~3), to 
the nearest neighboring galaxy having an apparent magnitude brighter than 
$B_T\,+\,1.5$ mag and a systemic velocity within $\upsilon_h\,\pm\,500$ \kms.  
These criteria select sizable objects, with luminosity ratios of at least 4:1, 
which are likely to be physically associated companions.  When no neighbor 
that satisfies the above criteria is found, $\Delta \theta$ is given as a lower
limit.  Following \citet{Bournaud05}, we also calculate the tidal 
parameter 

\begin{equation}
t_p \equiv \log \left\{ \sum_{i} \frac{M_i}{M_0} 
\left(\frac{R_0}{D_i}\right)^{3} \right\} ,
\end{equation}

\noindent 
where $M_0$ and $R_0$ are the mass and size of the galaxy in question, $M_i$ 
and $D_i$ are the mass and projected separation of neighbor $i$, and the 
summation is performed over all neighbors with systemic velocities within 
$\upsilon_h\,\pm\,500$ \kms\ in a region with a radius of 750 kpc.  For 
simplicity, we assume that all galaxies have the same mass-to-light ratio and 
adopt the Johnson $B$ band as the reference filter.  To the extent 
possible, we convert photometry listed in other bandpasses to the $B$ band, 
assuming colors typical of an Sbc galaxy (\citealt{Fukugita95, Peletier98}).
An uncertainty of up to 0.3 mag, which encompasses the 
extreme range of plausible $K$ corrections for different Hubble types, 
introduces an error of 0.12 dex in $t_p$.  We set $R_0 = 0.5 D_{25}$. Lastly, 
we indicate whether the galaxy belongs to the field or to a known galaxy group 
or cluster.  The main cluster in the southern sky pertinent to CGS is Fornax, 
and we draw our membership identifications from \citet{Ferguson89} and 
\citet{Ferguson90}.  Most of the group assignments come from the 2MASS-based 
group catalog of \citet{Crook07}.

The last two columns of Table~7 summarize the neutral hydrogen content. 
The \hi\  (21~cm) flux, in magnitude units, is defined such that $m_{21}^c = 
-2.5 \log f + 17.40$, where the integrated line flux $f$ is in units of 
Jy~\kms.  We adopted the correction for self-absorption as given in 
HyperLeda.  In the optically thin limit, the \hi\ mass is given by

\begin{equation}
M_{{\rm H~{I}}} = 2.36\times10^5 \ D_{L}^2\ f \, \, \, \, M_\odot,
\end{equation}

\noindent
where $D_{L}$ is the luminosity distance expressed in Mpc (\citealt{Roberts62}).  
We list the \hi\ mass normalized to the total $B$-band luminosity, using
magnitudes from Table~1, corrected for Galactic extinction.

\section{Summary}

The CGS is a long-term effort to investigate 
the detailed physical properties of a magnitude-limited ($B_T < 12.9$ mag), 
statistically complete sample of 605 galaxies in the southern hemisphere 
($\delta < 0$\deg).  The present-day constitution of a galaxy encodes rich 
clues to its formation mechanism and evolutionary pathway.  CGS aims to secure 
the necessary data to quantify the main structural components, stellar 
content, kinematics, and level of nuclear activity in a large, representative 
set of local galaxies spanning a wide range of morphological type, mass, and 
environment.  

This paper, the first in a series, gives a broad overview of the survey, 
defines the sample selection, and describes the optical imaging component of 
the program.  Over the course of 69 nights, we collected more than 6000 
\emph{BVRI} science images using the du~Pont 2.5 m telescope at Las Campanas 
Observatory.  The image quality is generally quite good: half of the images 
were taken under sub-arcsecond seeing conditions.  The CCD camera has a 
field of view (8\farcm9$\times$8\farcm9) sufficient to yield reasonably 
accurate sky subtraction for most of the sample, allowing us to reach a median 
surface brightness limit of $\sim27.5$, 26.9, 26.4, and 25.3 mag~arcsec$^{-2}$ 
in the $B$, $V$, $R$, and $I$ bands, respectively.  Although only roughly half 
of the data were taken under photometric conditions, we are able to devise a 
calibration strategy to establish a photometric zero point for the 
rest of the survey.  

We apply post-processing steps to generate several data products from the 
images that will be useful for later science applications: (1) three-band 
color composites; (2) star-cleaned images that can be used as templates to 
simulate high-redshift observations; (3) stacked \emph{BVRI} images to enhance 
the surface brightness sensitivity; (4) structure maps to emphasize 
high-spatial frequency features such as spiral arms, dust lanes, and nuclear 
disks; and (5) color index maps to trace the spatial variation of stellar 
content and dust reddening.  An image atlas showcases these digital images for 
each galaxy.  To facilitate subsequent scientific analyses of the sample, we 
collect an extensive set of ancillary data, including optical isophotal and 
photometric parameters derived from CGS itself and published information on 
multiwavelength photometry, internal kinematics, environmental variables, and 
neutral hydrogen content.  We pay particular attention to ensuring the 
accuracy and homogeneity of these databases.

Paper~II (Li et al. 2011) presents the one-dimensional isophotal analysis for
the survey, including radial profiles of surface brightness, 
color, and geometric parameters.  Fourier decomposition of the isophotes yields 
quantitative measures of the strength of bars, spiral arms, and global 
asymmetry.  Our team is actively using the CGS data for a number of other 
applications, including investigations of disk truncations, lopsidedness, 
color gradients, pseudobulges, and the fundamental plane of spheroids.  We are 
applying GALFIT (\citealt{Peng02, Peng10}) to perform full two-dimensional 
decomposition of the galaxies to obtain robust structural parameters for 
bulges, bars, disks, and other photometrically distinct subcomponents, and to 
quantify non-axisymmetric structures such as spiral arms and tidal distortions.

To maximize the science return of the survey, we intend to make accessible to 
the community all the final, fully reduced, calibrated science images, 
their associated ancillary products (object masks, PSF images, star-cleaned 
images, etc.), as well as higher-level science derivatives resulting from the 
one-dimensional and two-dimensional analysis.  We will strive to release 
these data, as they become available, roughly within one year of their initial 
publication.  Please consult the project Web site 
({\tt http://cgs.obs.carnegiescience.edu}) for updates.

\acknowledgements
We thank the referee for a prompt and helpful review of this manuscript.
This work was supported by the Carnegie Institution for Science and by the 
UC Irvine School of Physical Sciences.  Additional support was provided by the 
UC Irvine School of Information and Computer Sciences through a collaboration 
with Eric Mjolsness (A.J.B.), the China Scholarship Council (Z.-Y.L.), and the 
Plaskett Fellowship of the Herzberg Institute of Astrophysics, National 
Research Council of Canada (C.Y.P.).  Z.-Y.L. is grateful to Professor X.-B. Wu 
of the Department of Astronomy in Peking University for his support and helpful 
suggestions on this project.  We thank the Staff of the Observatories 
for their generous allocation of telescope time during the course of this 
survey.  Wojtek Krzeminski provided assistance for some of the observing 
runs.  We made use of the following astronomical databases: NASA/IPAC 
Extragalactic Database (NED), which is operated by the Jet Propulsion 
Laboratory, California Institute of Technology, under contract with the 
National Aeronautics and Space Administration, SIMBAD, which is operated 
at the Centre de Donn\'ees Astronomiques de Strasbourg, France, and HyperLeda.


\clearpage

\begin{figure*}[t]
\centerline{\psfig{file=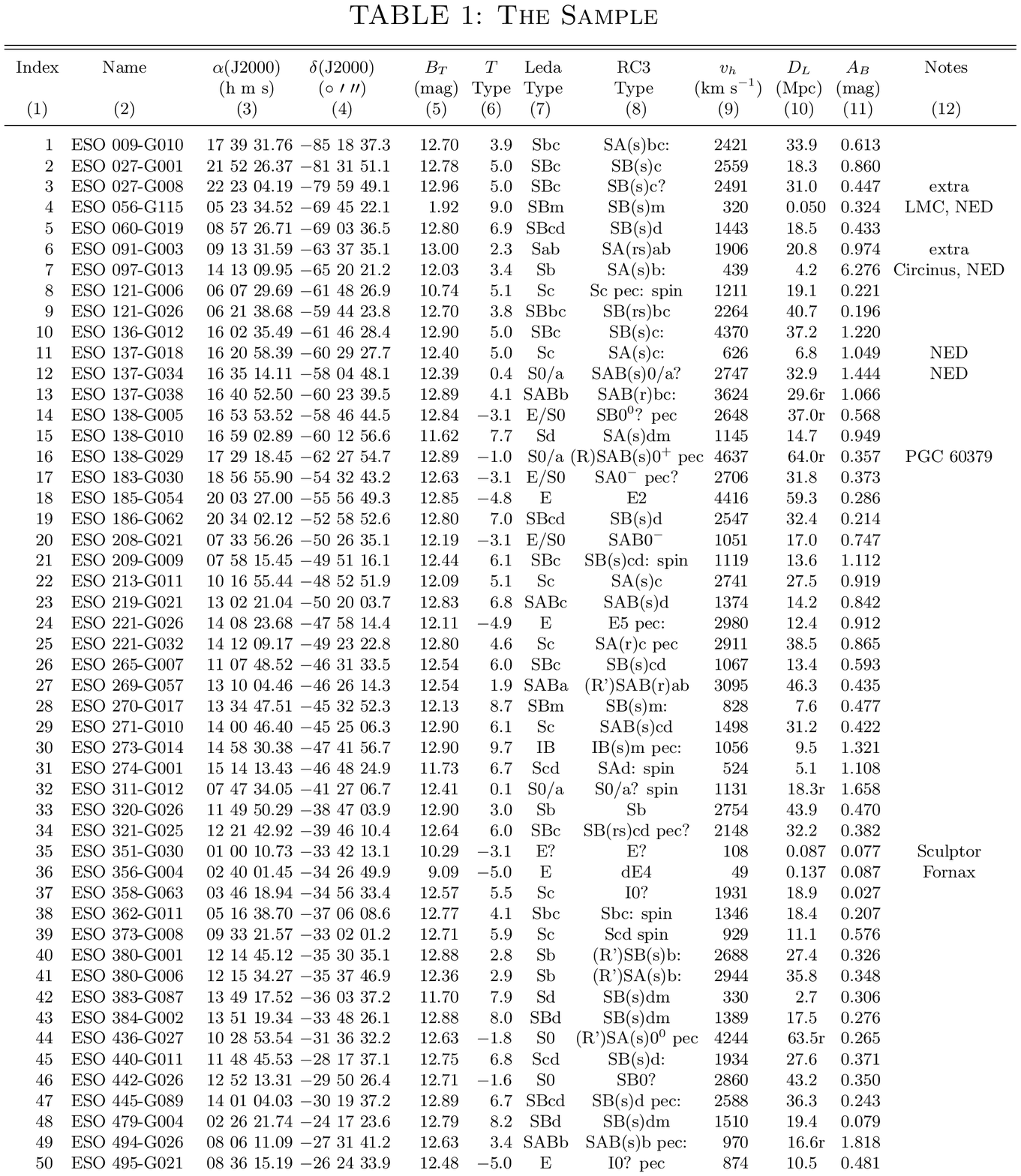,width=18.5cm,angle=0}}
\end{figure*}

\begin{figure*}[t]
\centerline{\psfig{file=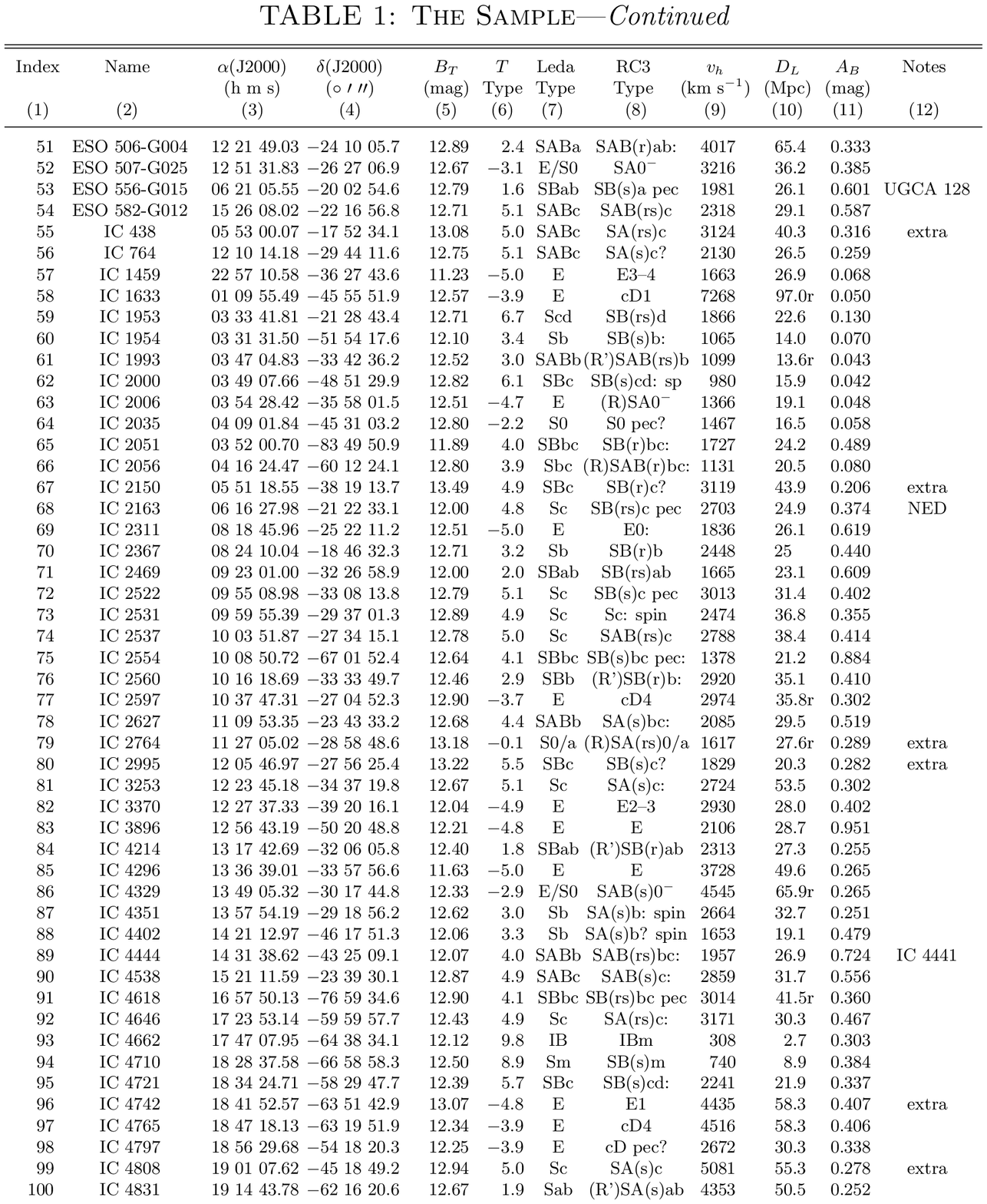,width=18.5cm,angle=0}}
\end{figure*}

\begin{figure*}[t]
\centerline{\psfig{file=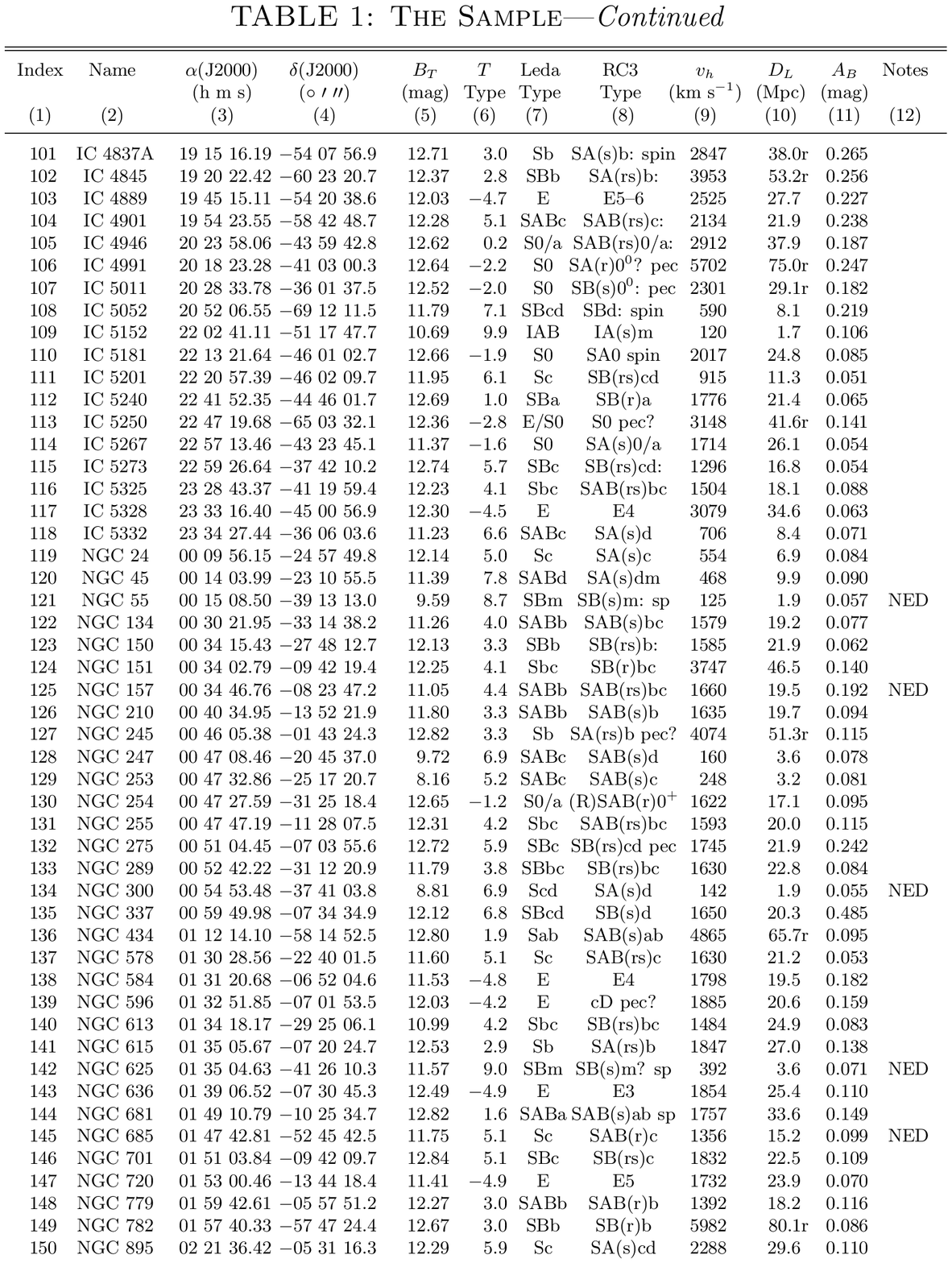,width=18.5cm,angle=0}}
\end{figure*}

\begin{figure*}[t]
\centerline{\psfig{file=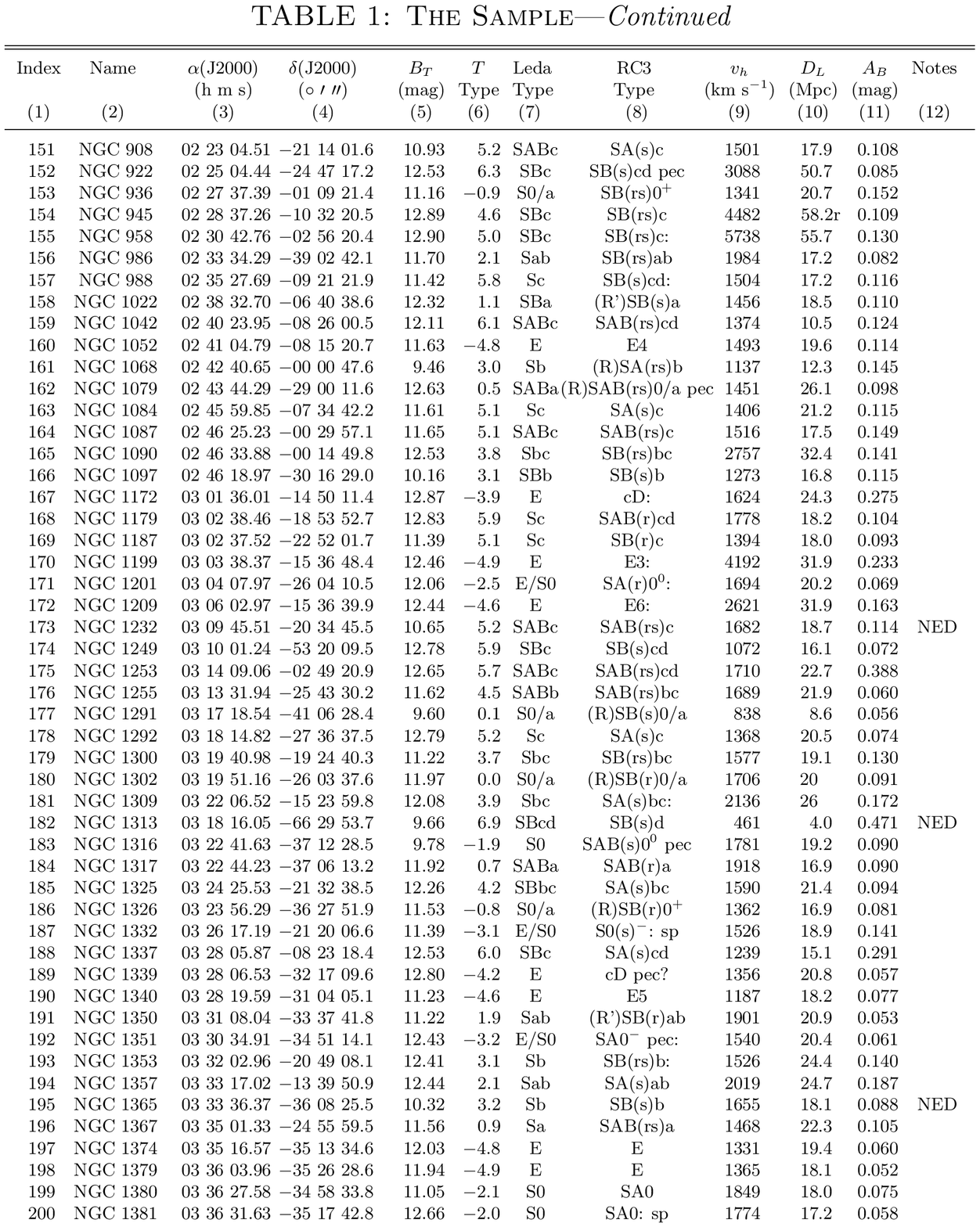,width=18.5cm,angle=0}}
\end{figure*}

\begin{figure*}[t]
\centerline{\psfig{file=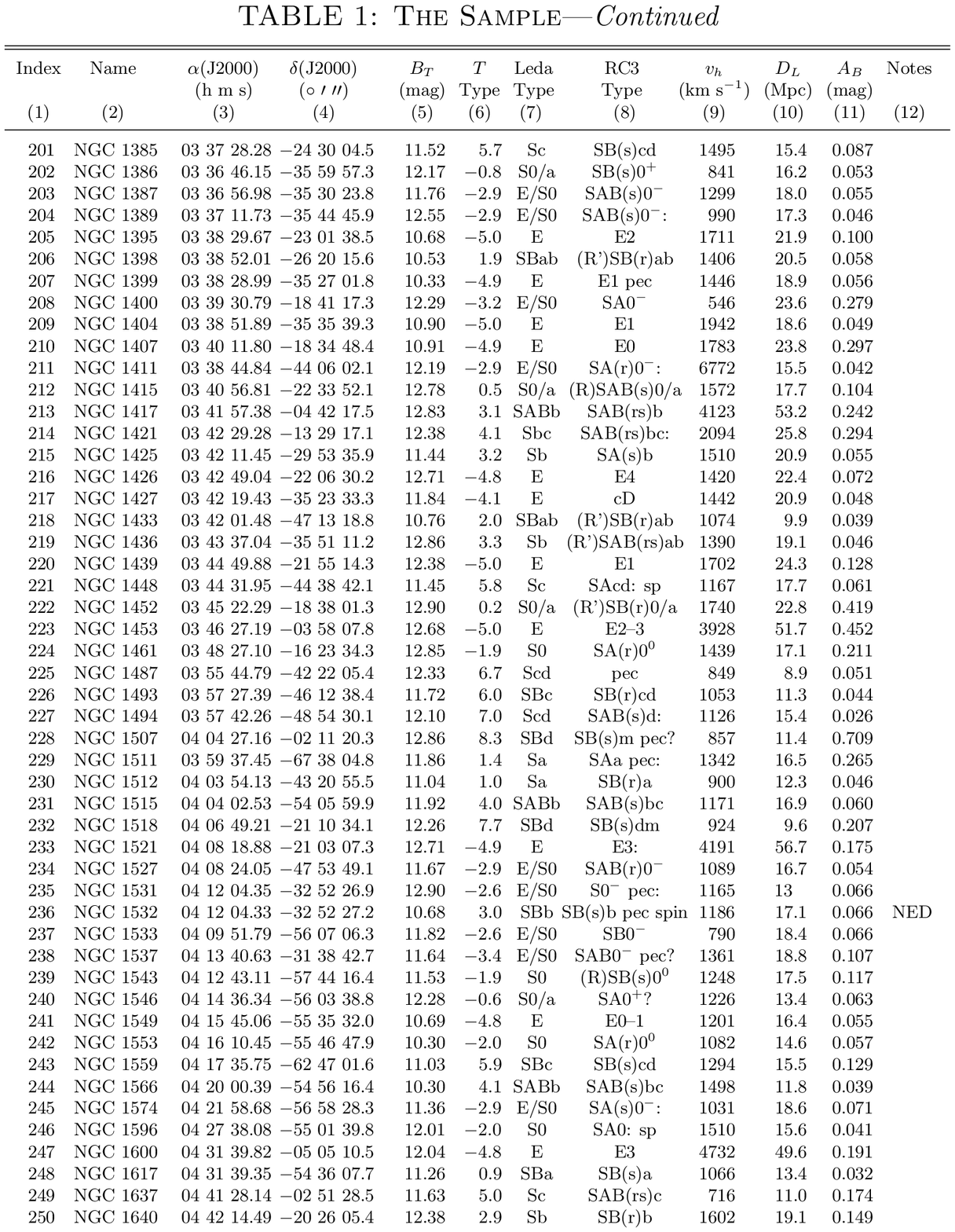,width=18.5cm,angle=0}}
\end{figure*}

\begin{figure*}[t]
\centerline{\psfig{file=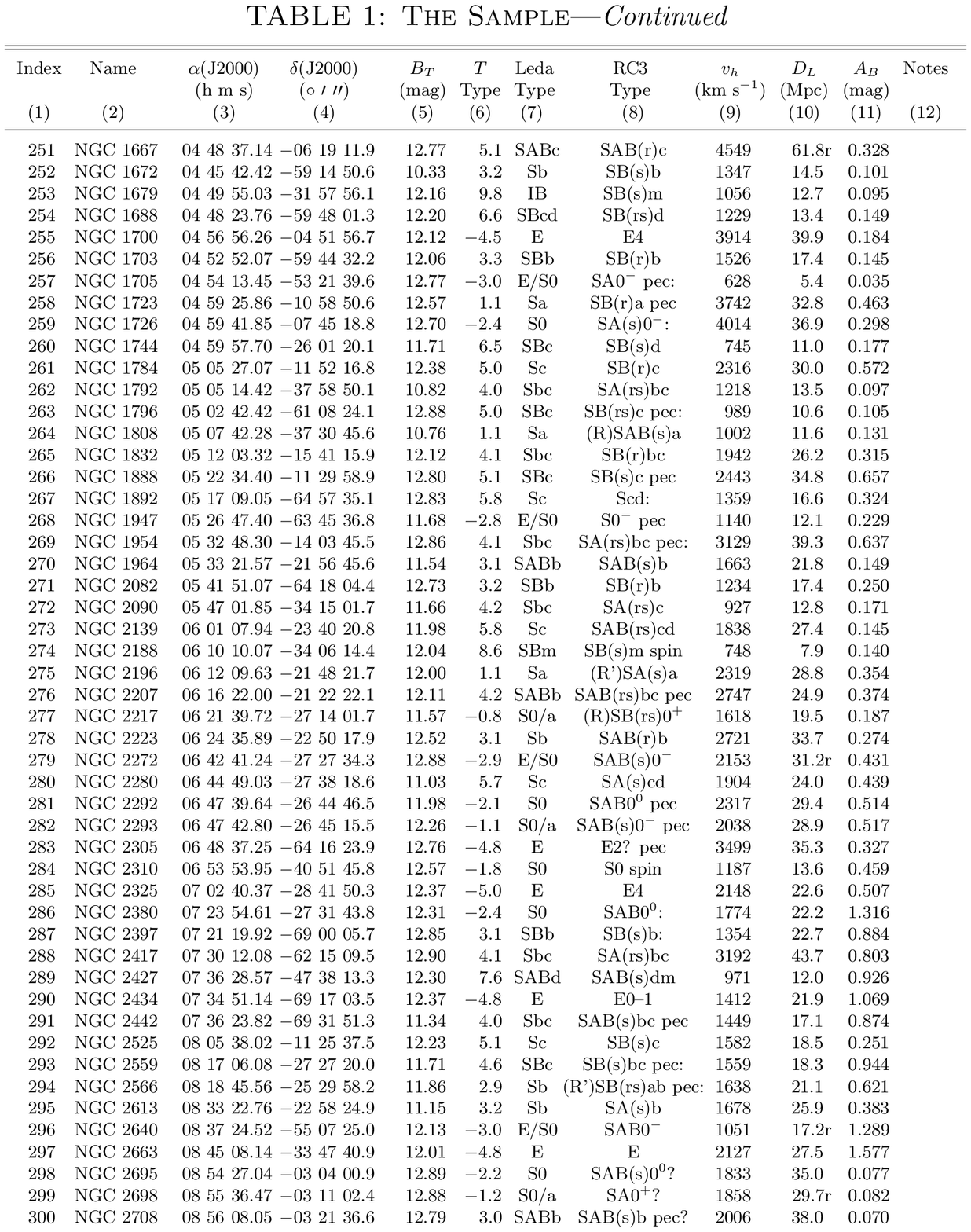,width=18.5cm,angle=0}}
\end{figure*}

\begin{figure*}[t]
\centerline{\psfig{file=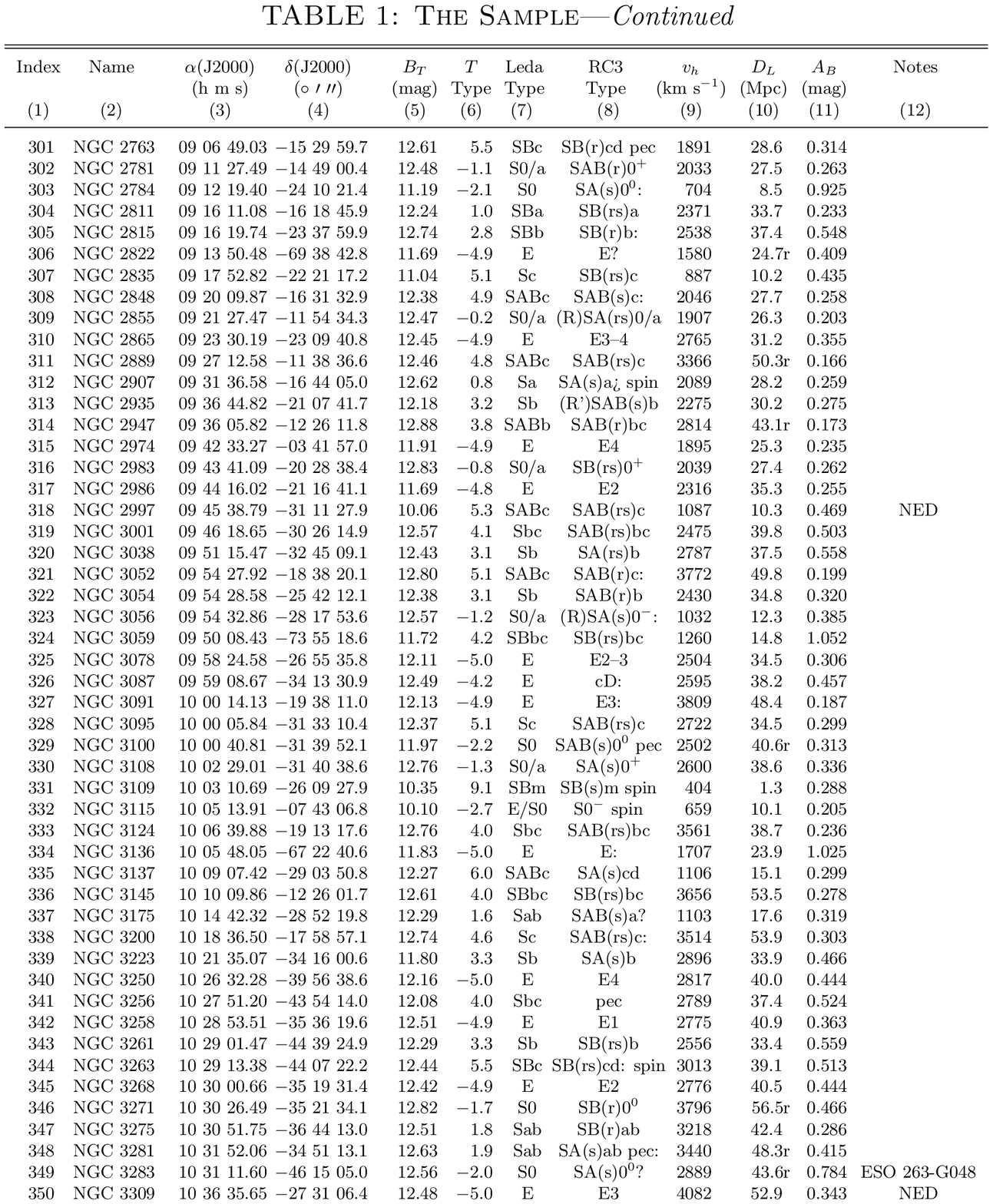,width=18.5cm,angle=0}}
\end{figure*}

\begin{figure*}[t]
\centerline{\psfig{file=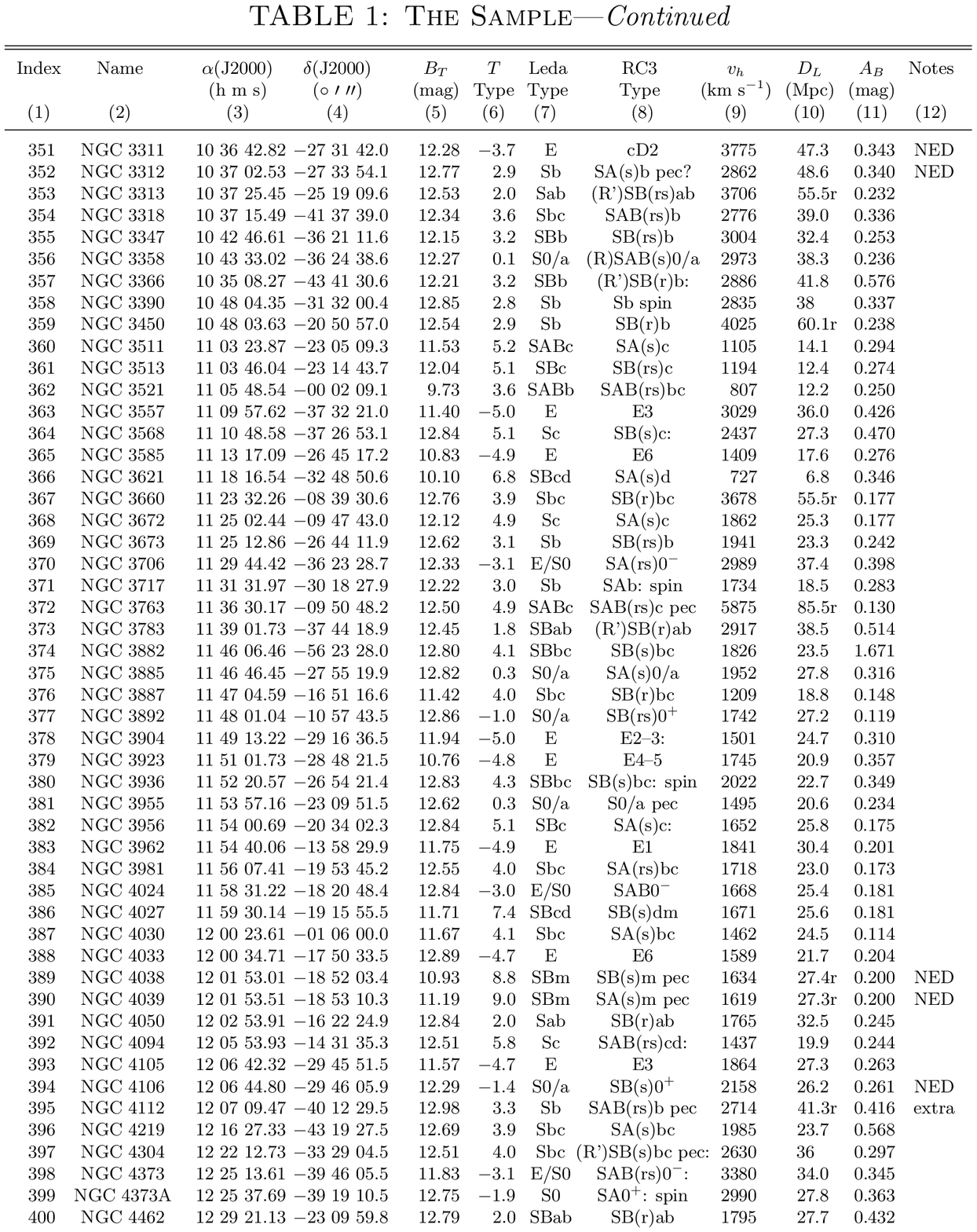,width=18.5cm,angle=0}}
\end{figure*}

\begin{figure*}[t]
\centerline{\psfig{file=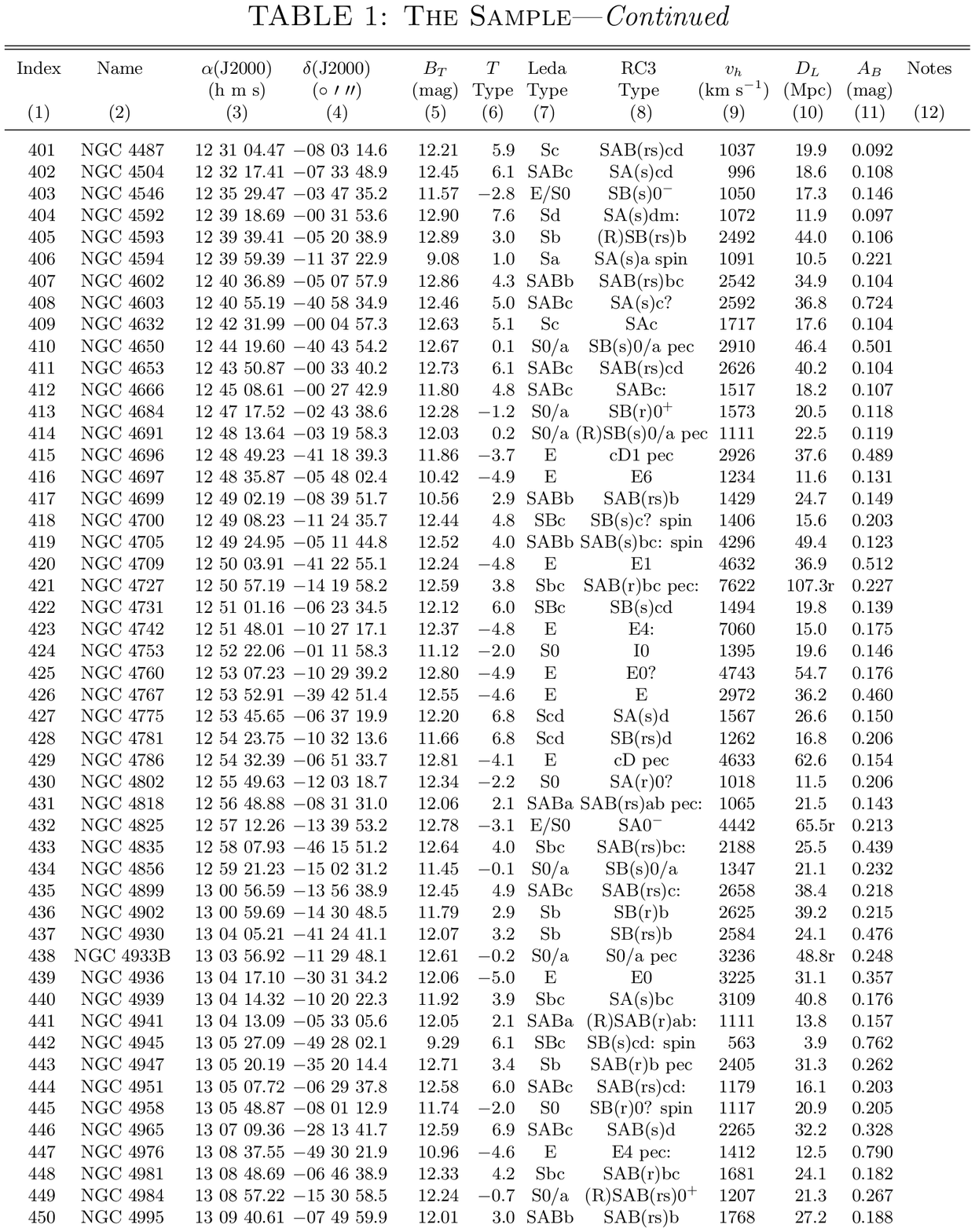,width=18.5cm,angle=0}}
\end{figure*}

\begin{figure*}[t]
\centerline{\psfig{file=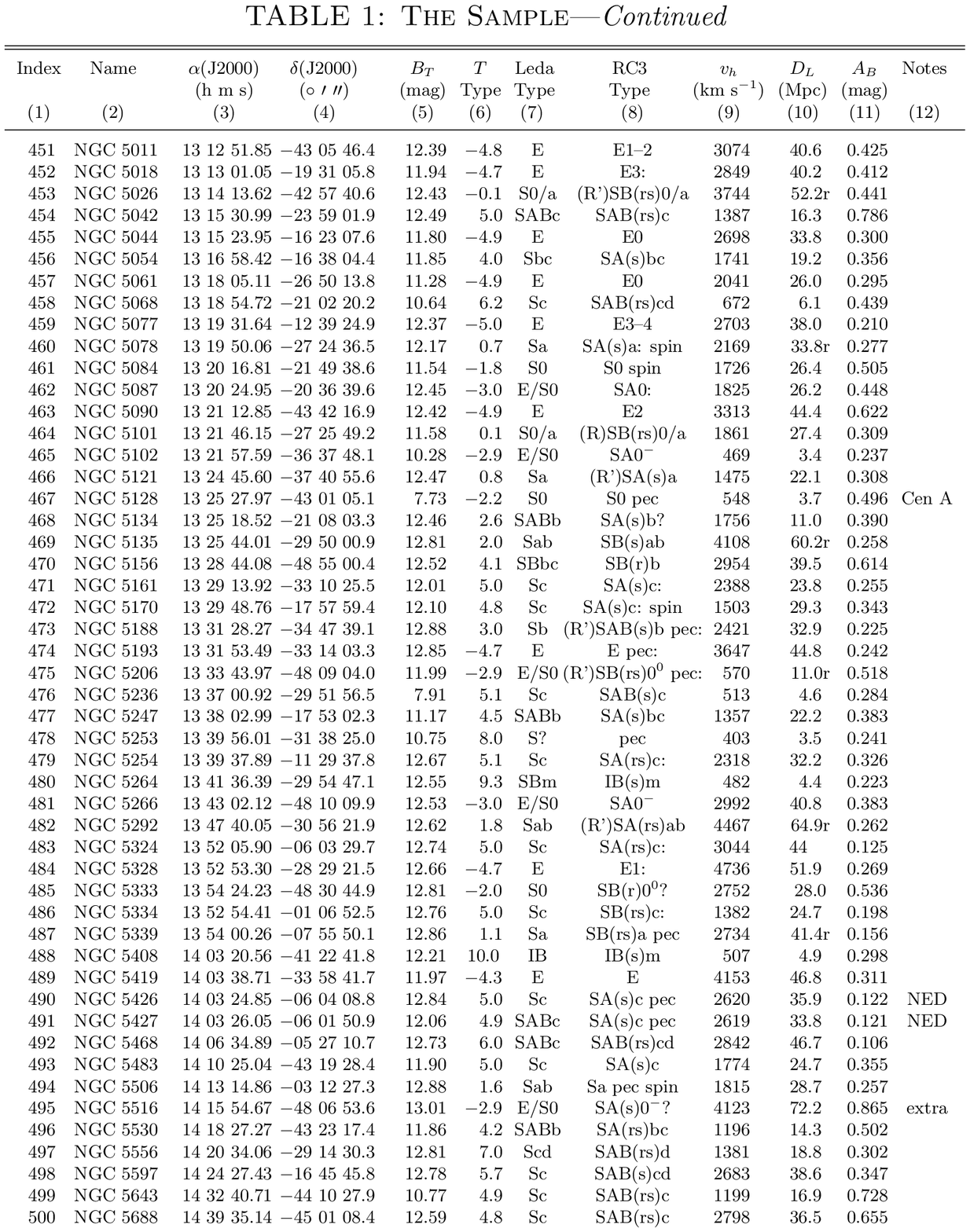,width=18.5cm,angle=0}}
\end{figure*}

\begin{figure*}[t]
\centerline{\psfig{file=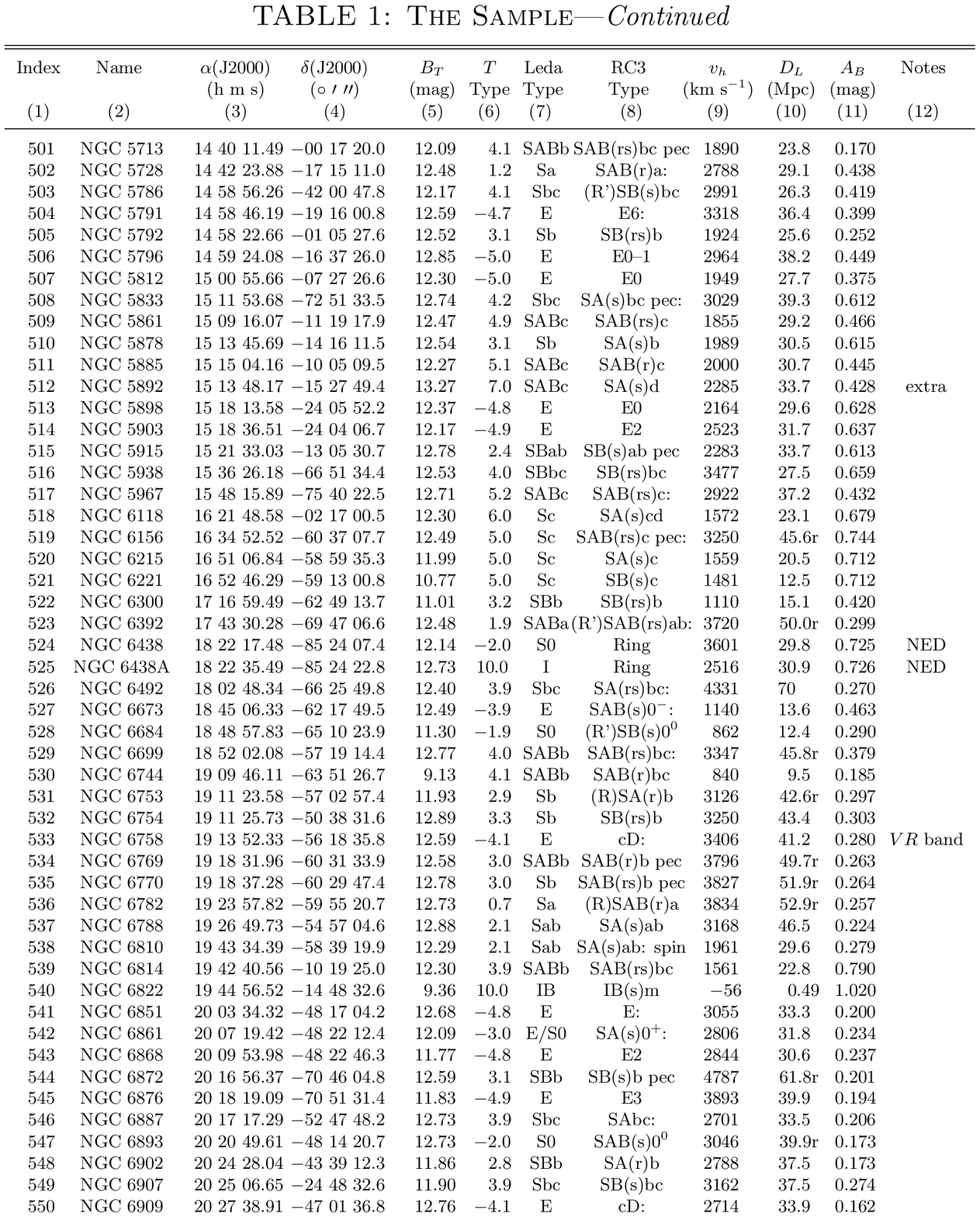,width=18.5cm,angle=0}}
\end{figure*}

\begin{figure*}[t]
\centerline{\psfig{file=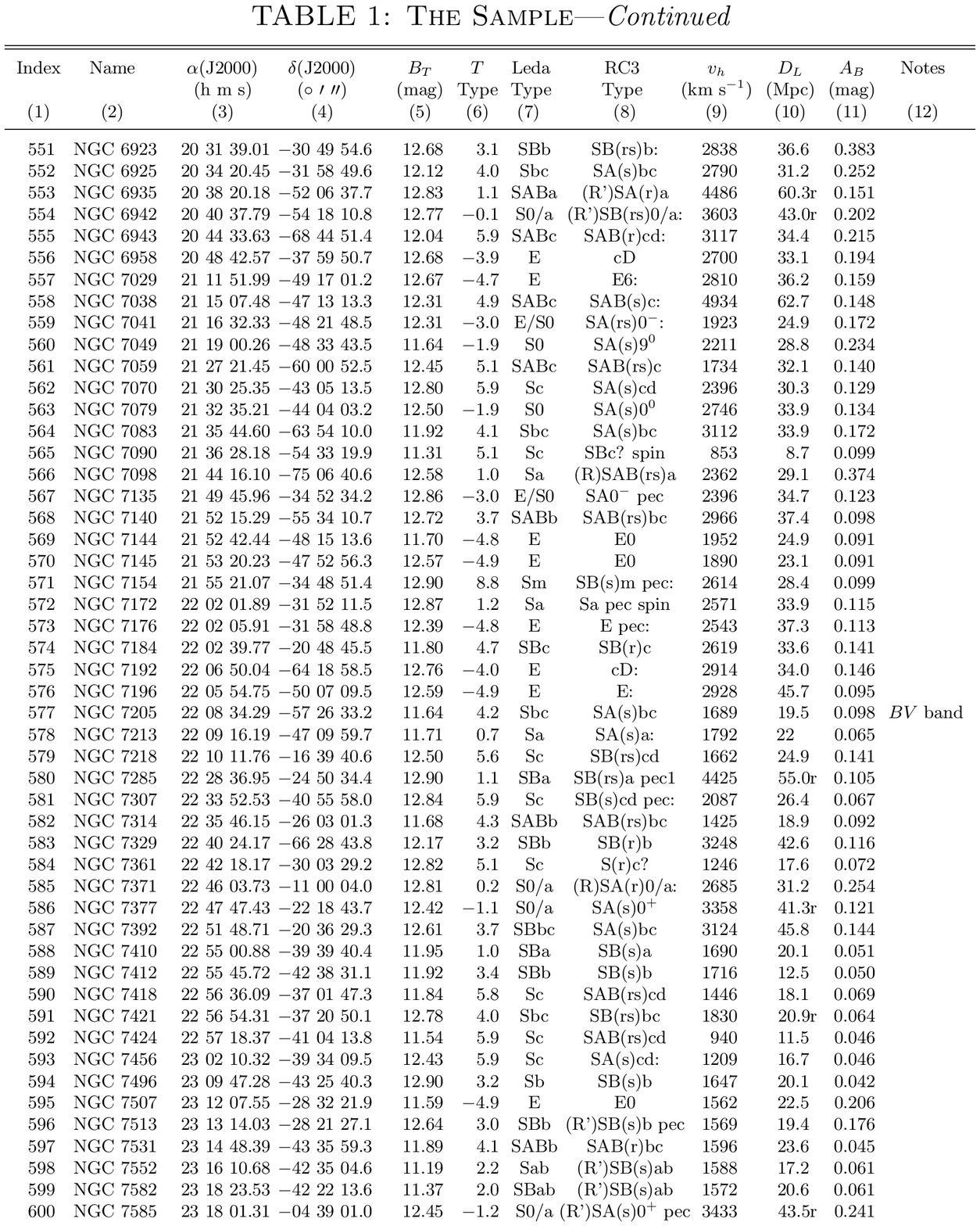,width=18.5cm,angle=0}}
\end{figure*}

\begin{figure*}[t]
\centerline{\psfig{file=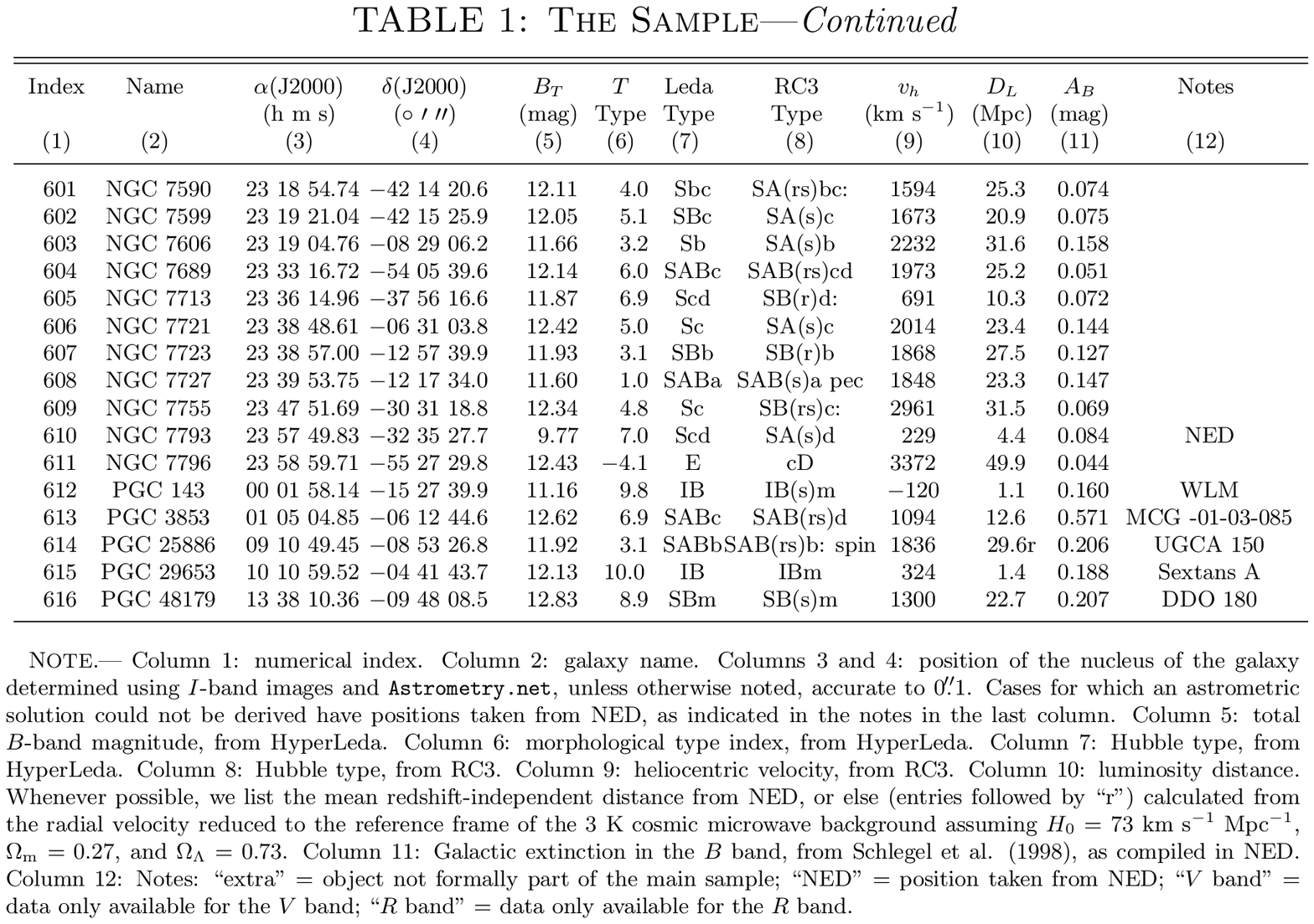,width=18.5cm,angle=0}}
\end{figure*}
\clearpage

\begin{figure*}[t]
\centerline{\psfig{file=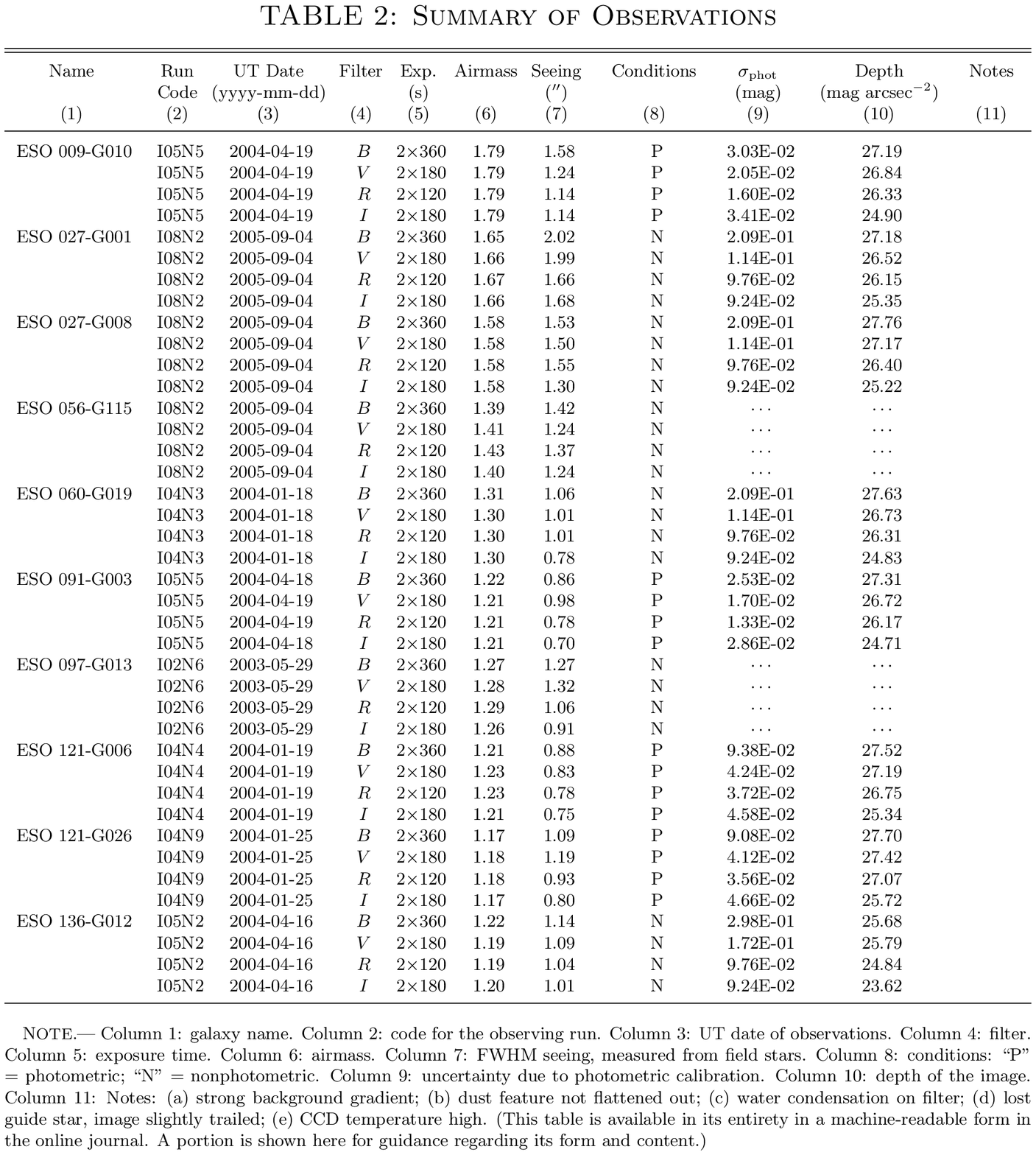,width=18.5cm,angle=0}}
\end{figure*}
\clearpage

\begin{figure*}[t]
\centerline{\psfig{file=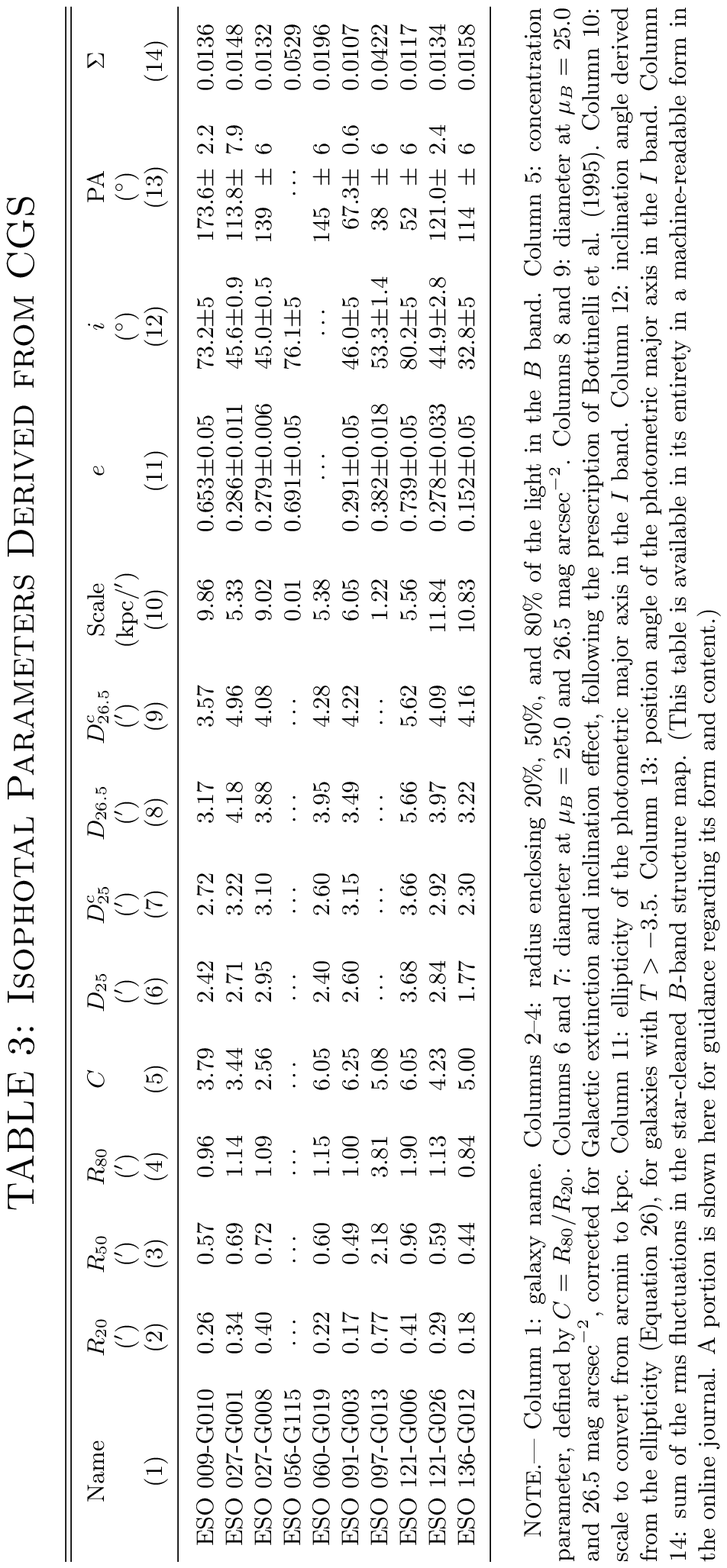,height=10.5in,angle=180}}
\end{figure*}
\clearpage

\begin{figure*}[t]
\centerline{\psfig{file=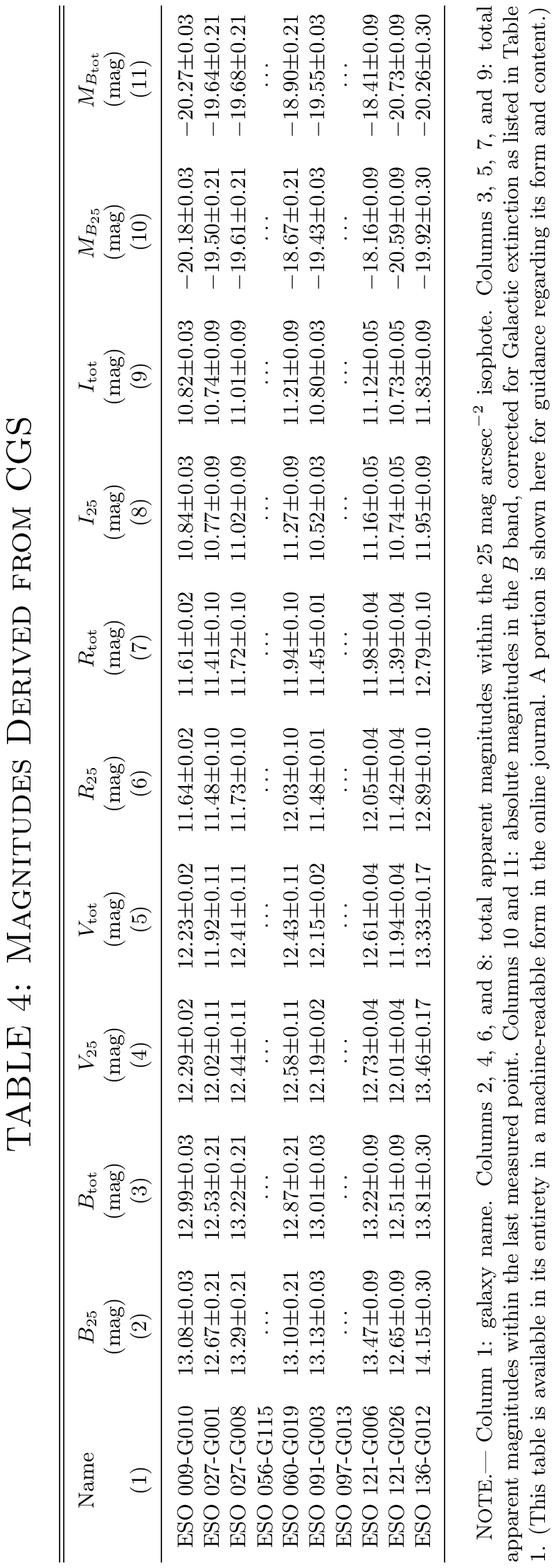,height=10.5in,angle=180}}
\end{figure*}
\clearpage

\begin{figure*}[t]
\centerline{\psfig{file=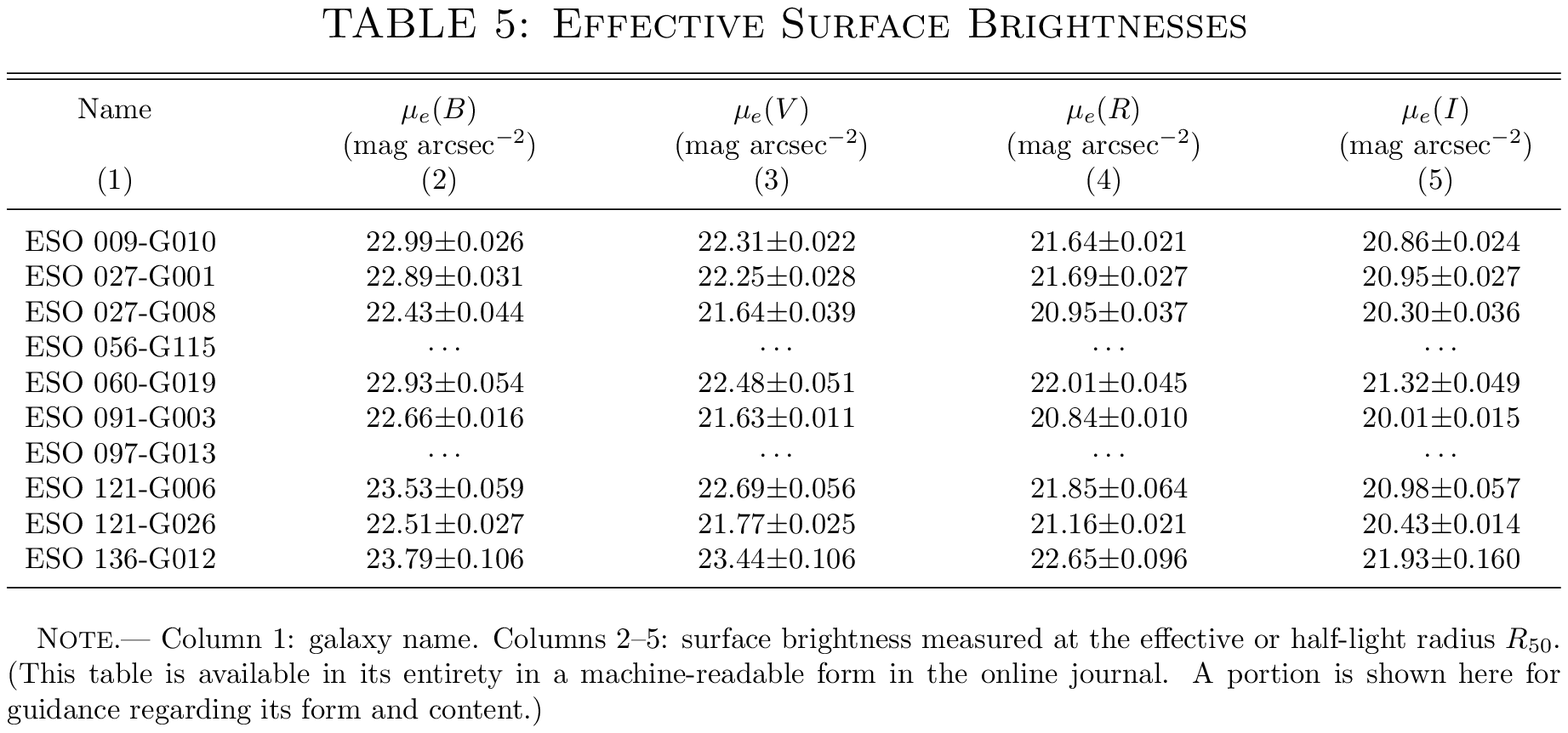,width=18.5cm,angle=0}}
\end{figure*}
\clearpage

\begin{figure*}[t]
\centerline{\psfig{file=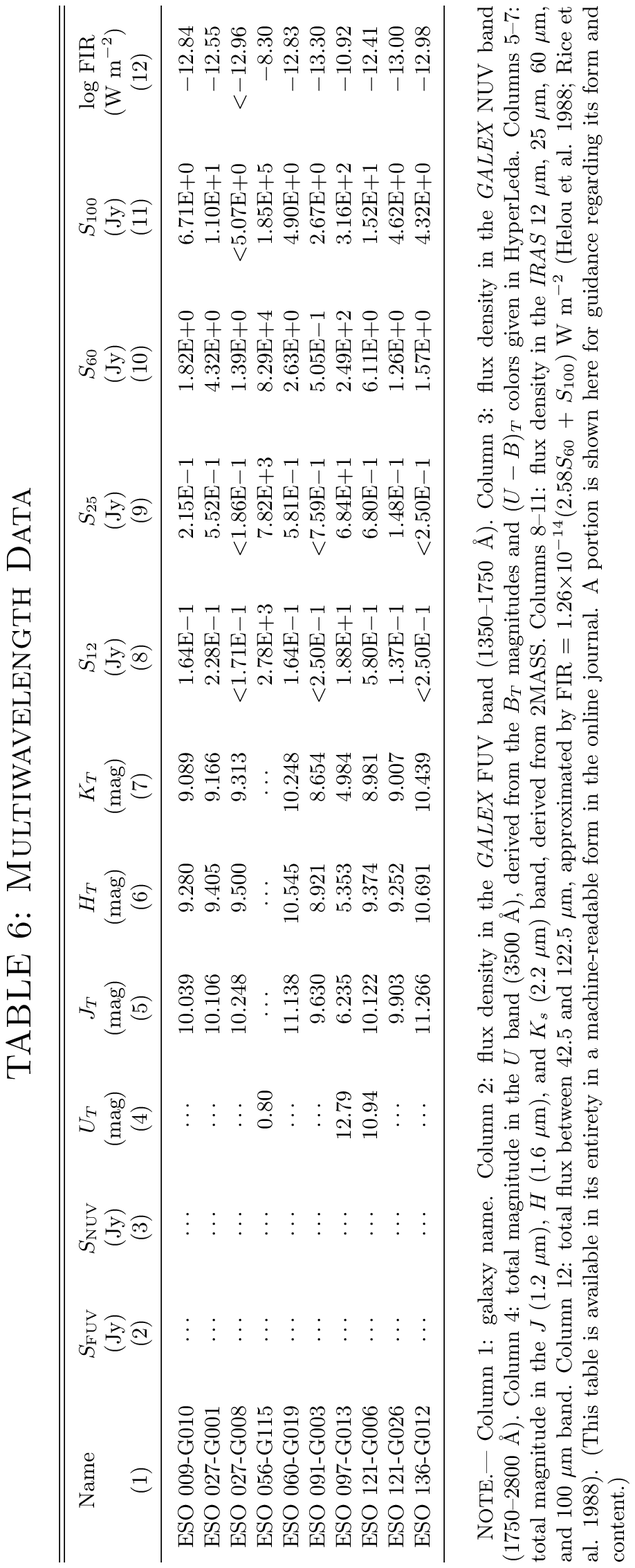,height=10.5in,angle=180}}
\end{figure*}
\clearpage

\begin{figure*}[t]
\centerline{\psfig{file=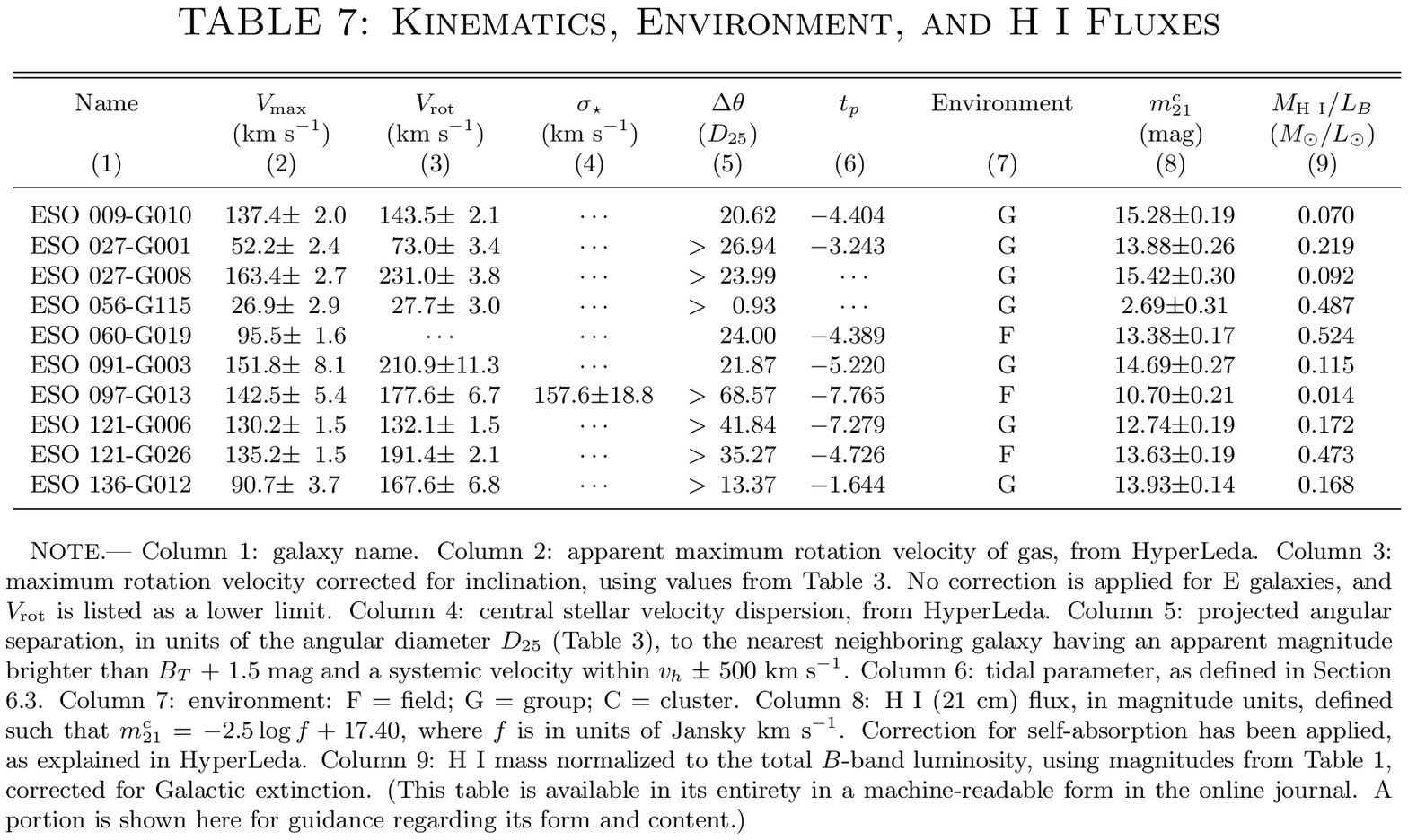,width=18.5cm,angle=0}}
\end{figure*}

\clearpage
\appendix

\section{Image Atlas}

Figures~7.1--7.616 present the image atlas for the 605 galaxies in CGS; we also 
include the 11 extra galaxies that are not part of the formal sample.  One
full-page figure is devoted to each galaxy, ordered sequentially following the
numerical indices listed in Table~1.  The six panels of each figure show the
composite color image, the star-cleaned composite color image, the \emph{BVRI}
stacked image, the structure map of the star-cleaned $B$-band image, and the 
$B-R$ and $R-I$ color index maps.  Darker regions on the index maps correspond 
to redder colors.  Each image is scaled to a dimension of $1.5\, D_{25}$, with 
north up and east to the left.  We use an arcsinh stretch for the color 
composite, star-cleaned, and stacked images, while the structure and color 
index maps are shown on a linear stretch.  We only display three sample pages 
for illustration (Figures 7(a)--7(c)); the full set of figures is available in 
the electronic version of the paper, as well as on the project Web site 
{\tt http://cgs.obs.carnegiescience.edu}.

\begin{figure*}[t]
\figurenum{7(a)}
\centerline{\psfig{file=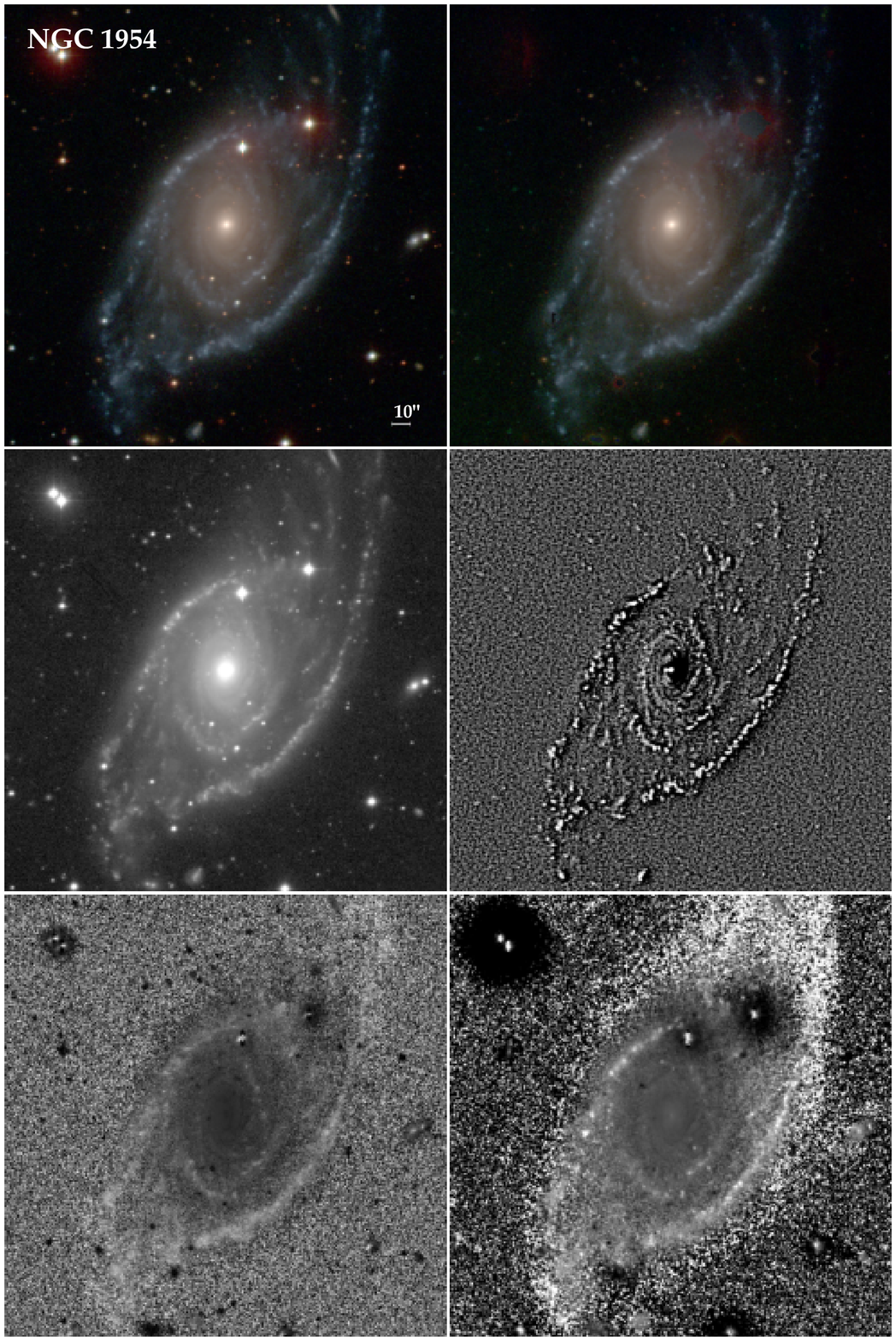,width=16.5cm,angle=0}}
\figcaption[figA269.ps]{
Sample page from the image atlas, illustrating NGC 1954.  The panels show, from
top to bottom, left to right: the composite color image, the star-cleaned
composite color image, the \emph{BVRI} stacked image, the structure map of the
star-cleaned $B$-band image, and the $B-R$ and $R-I$ color index maps.  Darker
regions on the index maps correspond to redder colors.  Each image is scaled
to a dimension of $1.5\, D_{25}$, with north up and east to the left.  The
color composite, star-cleaned, and stacked images are shown on an arcsinh
stretch, and the rest are shown on a linear stretch.
\label{figA269}}
\end{figure*}

\begin{figure*}[t]
\figurenum{7(b)}
\centerline{\psfig{file=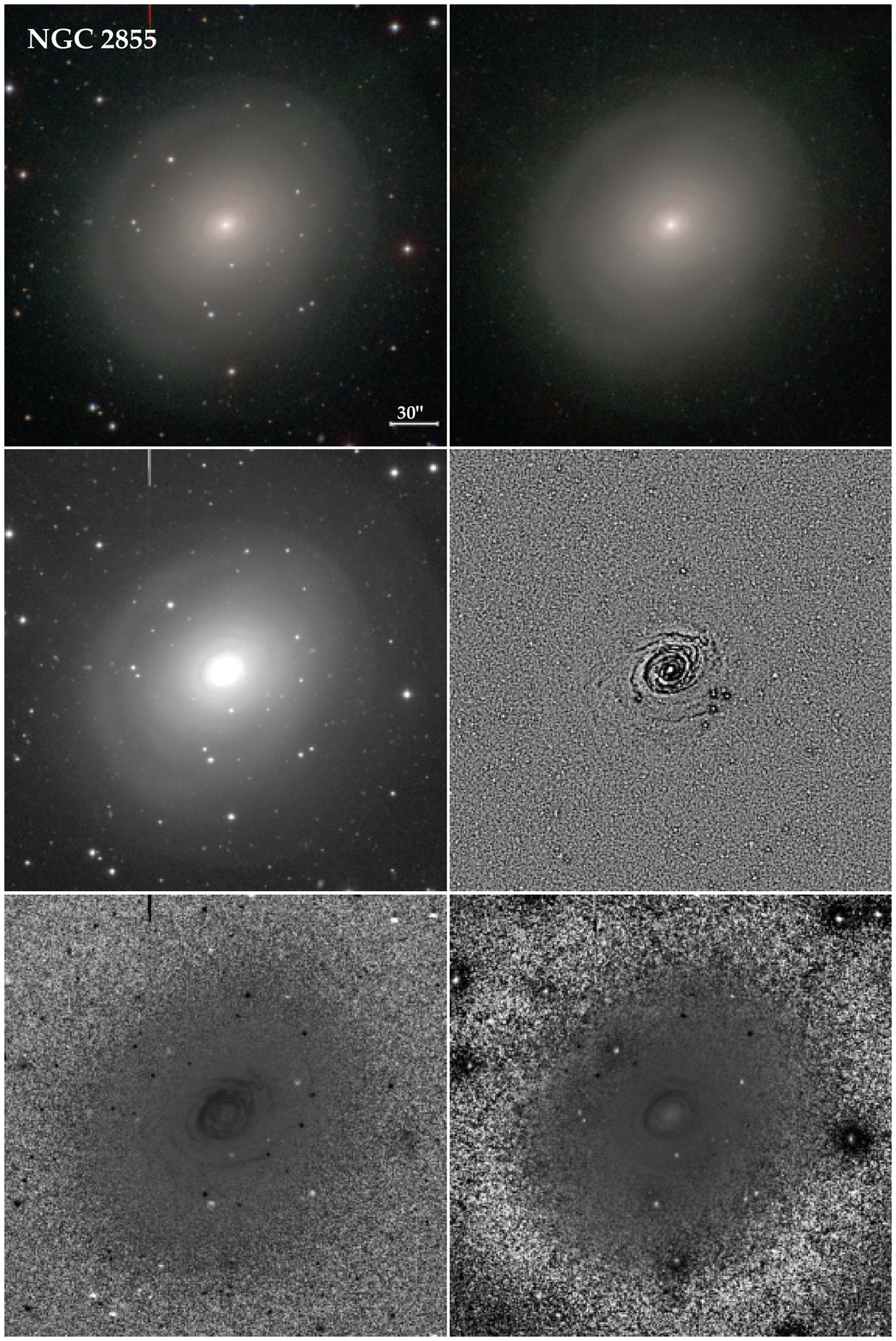,width=16.5cm,angle=0}}
\figcaption[figA309.ps]{
Sample page from the image atlas, illustrating NGC 2855.  The panels show, from
top to bottom, left to right: the composite color image, the star-cleaned
composite color image, the \emph{BVRI} stacked image, the structure map of the 
star-cleaned $B$-band image, and the $B-R$ and $R-I$ color index maps.  Darker 
regions on the index maps correspond to redder colors.  Each image is scaled 
to a dimension of $1.5\, D_{25}$, with north up and east to the left.  The 
color composite, star-cleaned, and stacked images are shown on an arcsinh 
stretch, and the rest are shown on a linear stretch.
\label{figA309}}
\end{figure*}

\begin{figure*}[t]
\figurenum{7(c)}
\centerline{\psfig{file=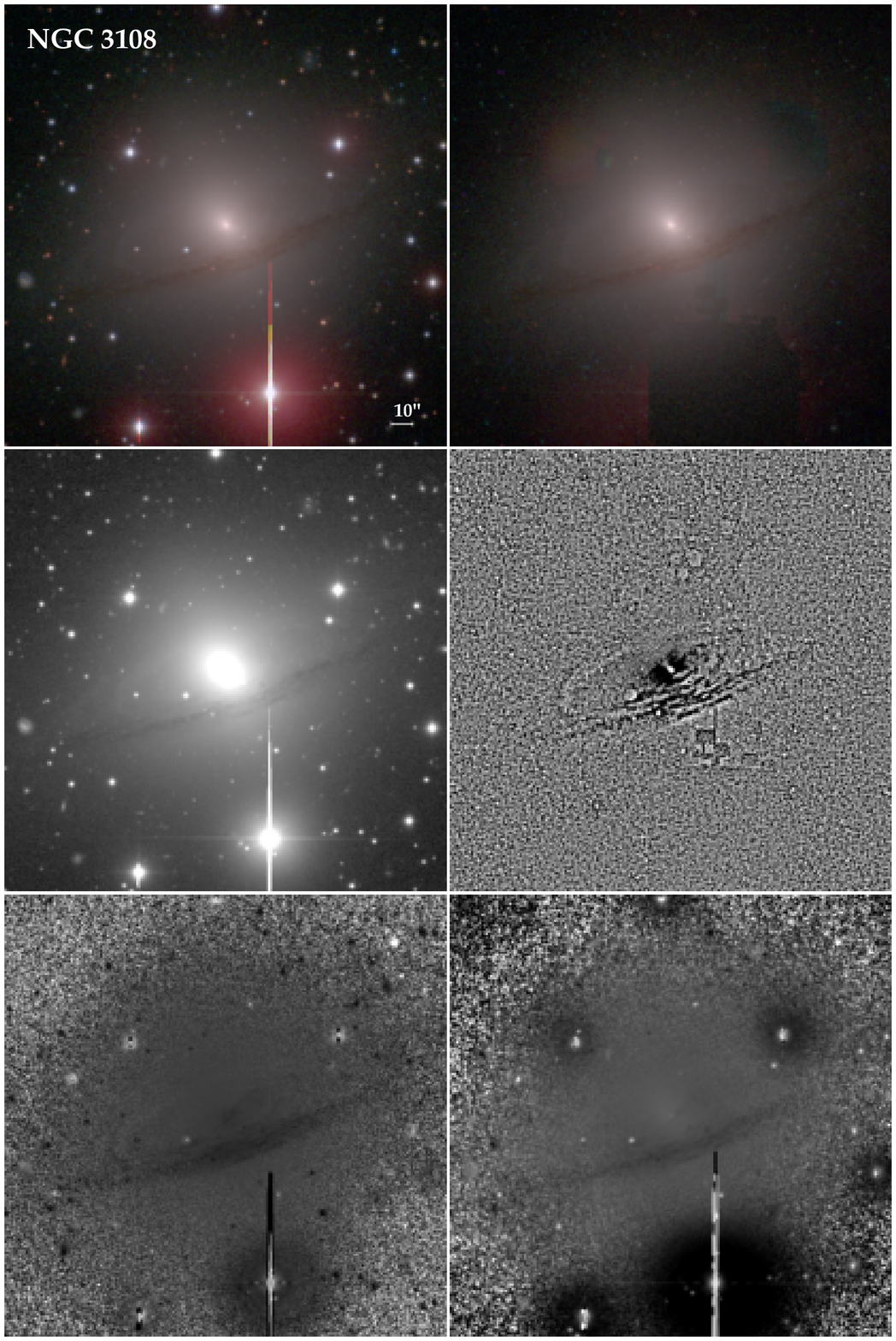,width=16.5cm,angle=0}}
\figcaption[figA330.ps]{
Sample page from the image atlas, illustrating NGC 3108.  The panels show, from
top to bottom, left to right: the composite color image, the star-cleaned
composite color image, the \emph{BVRI} stacked image, the structure map of the 
star-cleaned $B$-band image, and the $B-R$ and $R-I$ color index maps.  Darker 
regions on the index maps correspond to redder colors.  Each image is scaled 
to a dimension of $1.5\, D_{25}$, with north up and east to the left.  The 
color composite, star-cleaned, and stacked images are shown on an arcsinh 
stretch, and the rest are shown on a linear stretch.
\label{figA330}}
\end{figure*}

\clearpage

\end{document}